\documentclass[twocolumn,times]{aastex63}

\usepackage{amsmath}
\usepackage{soul}

\shorttitle{X-ray emission in Green Peas}
\shortauthors{Mainak Singha et al.}
\graphicspath{{./}{figures/}}
\def\hide#1{}

\begin{document}
\title{\textit{Chandra} detects low-luminosity AGN  with $M_\mathrm{BH}=10^{4}-10^{6}~M_\mathrm{\odot}$ in nearby ($z<0.5$), dwarf and star-forming galaxies} 
\author[0000-0001-5687-1516]{Mainak~Singha}

\affiliation{Astrophysics Science Division, NASA, Goddard Space Flight Center, Greenbelt, MD 20771, USA}

\affiliation{Department of Physics, The Catholic University of America, Washington, DC 20064, USA}

\affiliation{Center for Research and Exploration in Space Science and Technology, NASA, Goddard Space Flight Center, Greenbelt, MD 20771, USA}

\author{Julissa~Sarmiento}
\affiliation{Department of Physics and Astronomy, University of Pittsburgh, Pittsburgh, PA 15260, USA}

\author[0000-0002-9226-5350]{Sangeeta~Malhotra}
\affiliation{Astrophysics Science Division, NASA, Goddard Space Flight Center, Greenbelt, MD 20771, USA}

\author{James~E.~Rhoads} 
\affiliation{Astrophysics Science Division, NASA, Goddard Space Flight Center, Greenbelt, MD 20771, USA}

\author[0000-0003-3466-035X]{L.~Y.~Aaron~Yung}
\affiliation{Space Telescope Science Institute, 3700 San Martin Drive, Baltimore, MD 21218, USA}

\author{Junxian~Wang}
\affiliation{CAS Key Laboratory for Research in Galaxies and Cosmology, Department of Astronomy, University of Science and Technology of China, Hefei 230026, China}

\author[0000-0002-9634-2923]{Zhen-Ya~Zheng}
\affiliation{Key Laboratory for Research in Galaxies and Cosmology, Shanghai Astronomical Observatory, Chinese Academy of Sciences, 80 Nandan Road, Shanghai 200030, China}
\affiliation{School of Astronomy and Space Sciences, University of Chinese Academy of Sciences, Beijing 100049, China}

\author{Ruqiu~Lin}
\affiliation{Key Laboratory for Research in Galaxies and Cosmology, Shanghai Astronomical Observatory, Chinese Academy of Sciences, 80 Nandan Road, Shanghai 200030, China}
\affiliation{School of Astronomy and Space Sciences, University of Chinese Academy of Sciences, Beijing 100049, China}

\author[0000-0001-5687-1516]{Keunho~Kim}
\affiliation{IPAC, California Institute of Technology, 1200 E. California Boulevard, Pasadena, CA 91125, USA}

\author{Jialai~Kang}
\affiliation{University of Science and Technology of China, Hefei 230026, People’s Republic of China}

\author{Santosh~Harish}
\affiliation{Laboratory for Multiwavelength Astrophysics, School of Physics and Astronomy, Rochester Institute of Technology, 84 Lomb Memorial
Drive, Rochester, NY 14623, USA}

\begin{abstract}
We searched the Chandra and XMM archives for observations of 900 green pea galaxies to seek signatures of AGN. Green peas are low-mass galaxies with prominent emission lines, and have remarkably similar properties (such as size, star formation rate) compared to redshift, $z>7$ dwarf galaxies.  Out of the 29 observations found in the archives, 9 show detections in X-rays at $S/N>3$. The upper limits of others are also pretty similar. 
The 2-10 keV X-ray luminosity for the 9 sources exceeds $10^{40}~\mathrm{erg~s}^{-1}$, and 2 sources show 2-10 keV X-ray luminosity greater than $10^{41}~\mathrm{erg~s}^{-1}$, suggesting the presence of intermediate mass black holes (IMBH)/low-luminosity AGN (LLAGN), corresponding to BH mass, $M_\mathrm{BH}=100-10^6M_\mathrm{\odot}$. All of the X-ray detected sources (plus 6 additional sources) show He~II$\lambda4686$ emission and  a broad component of the H$\alpha$ emission line indicating the presence of winds. The line widths of the broad H$\alpha$ and He~II$\lambda4686$ emitting gas clouds are only weakly correlated ($R^{2}=0.15$), suggesting that He~II$\lambda4686$ emission is inconsistent with winds from super-Eddington accretors. However, the ratio of X-ray luminosity to star-formation rate is consistent with an anti-correlation with metallicity for 5 out of 9 sources with X-ray detection. Such anti-correlation suggests that ultraluminous X-ray sources could be the key contributors to the observed X-ray luminosity, which could be either: (i) super-Eddington accretors, or (ii) intermediate mass black holes (IMBH). The observed X-ray emission is at least an order of magnitude higher than what could be produced by Wolf-Rayet stars and fast radiative shocks from supernovae-driven winds. The X-ray luminosity in all 9 sources could, therefore, only be explained by black holes with masses exceeding 100~$M_\mathrm{\odot}$. Our results suggest that the X-ray emission in our sources requires a LLAGN. The line widths of broad H$\alpha$, if due to accretion onto massive BH, imply masses of $M_\mathrm{BH}=10^4-10^6M_\mathrm{\odot}$. Since Green Peas have been shown to be significant Lyman Continuum leakers, the presence of LLAGN in these galaxies would imply that that such AGN might have significantly contributed to the cosmic reionization.
\end{abstract}


\section{Introduction}\label{sec:intro}
A fundamental unsolved problem in observational cosmology galaxy formation is the role of accreting supermassive black holes (SMBH) or active galactic nuclei (AGN) in ionizing the universe during the cosmic reionization. AGN fuelled by the accreting energy are more efficient in producing hard-ionizing photons—such as He~II$\lambda4686$ \citep{Giallongo2015,Yung2021}. This leads to many studies propose the idea that AGN might have served as a significant source of ionizing photons, compared to starburst-driven processes \citep{Madau2015,Volonteri2016}. Whereas the general consensus is that the AGN number density is insufficiently high at elevated redshifts to accomplish complete reionization. Instead, it is widely believed that galactic ultraviolet (UV) sources play a key role in this process \citep{Robertson2015,Naidu2020, Yung2020b, Yung2020a}. Nevertheless, it is imperative that AGN can only dominate the reionization energy budget if there is an AGN abundance at $z>5$. 

Recent JWST observations have reported  an overdensity of AGN within the redshift range $z=4-12$. These AGN span a broad range in their Eddington ratios ($\lambda_\mathrm{Edd}=0.06-2$), bolometric luminosities ($L_\mathrm{bol}=10^{43}-10^{46}~\mathrm{erg~s}^{-1}$), and black hole masses ($M_\mathrm{BH}=10^{5}-10^{8}~M_\odot$) \citep{Harikane2023,Larson2023,Juodzbalis2023}. Studies employing JWST/NIRSpec observations propose a lower limit for unobscured AGN fractions between $z=6.5-12$, estimated at approximately $5\%$ \citep{Juodzbalis2023,Xu2023}. A recent JWST/ MIRI imaging study reported that around 25\% of galaxies in their sample at redshifts $z=3-5$ harbor heavily obscured AGN and composite AGN \citep{Yang2023}. This discovery suggests a threefold increase in the rate of black hole growth at these redshifts, exceeding expectations based on X-ray data. Such findings have substantial implications for our understanding of cosmic reionization, as accelerated black hole growth may imply an overdensity at these redshifts. Additionally, popular black hole growth models suggest that unobscured AGN is a direct result of the obscuring gas clouds being blown out by the AGN-driven superwinds \citep{Alexander2012}.  However, we need to keep in mind that at $z>5$: (1) AGN observations are relatively scarce, and (2) dwarf starburst galaxies dominate the overall galaxy population \citep{Robertson2015,Atek2024}. Therefore, local analogs of $z>5$ AGN host galaxies provide us with an attractive opportunity to better predict the presence of the first black holes, understand their properties and growth scenarios, and quantify their potential role in cosmic reionization.

Nearby ($z\sim0.3$), low-metallicity dwarf galaxies also known as green peas \citep{Cardamone2009} have garnered significant attention due to their remarkable similarities with $z>5$ Ly$\alpha$ galaxies as seen from recent JWST spectroscopic observations \citep{Rhoads2023}. Such similarities include having subsolar metallicities, [O~III]5007\AA\ rest-frame equivalent width exceeding $500$\AA\, half-light radius, $r_\mathrm{50}<0.3$ kpc, and an interacting nature \citep{Izotov2011a,Malhotra2012,Henry2015,Yang2016}.

In the past two decades, X-ray observations have extensively searched for AGN in Lyman$\alpha$ emitters at redshifts $z>2$ \citep{Malhotra2003,Basu-Zych2004,Wang2004b,Yang2009,Zheng2012,Calhau2020}. A commonly used indicator for AGN identification is the 2-10 keV X-ray luminosity, $L_\mathrm{2-10~keV}>10^{42}~\mathrm{erg~s}^{-1}$ \citep{Nandra2002,Haines2012,Birchall2022}. However, solely focusing on high-luminosity AGN might overlook those with lower luminosities ($L_\mathrm{2-10~keV}<10^{42}~\mathrm{erg~s}^{-1}$). For instance, our Milky Way's supermassive black hole, Sgr A*, has a quiet-state X-ray luminosity of $L_\mathrm{2-10~keV}\sim10^{33-35}~\mathrm{erg~s}^{-1}$ \citep{Sabha2010}. Recent IXPE observations have shown flaring events boosting its X-ray luminosity to $L_\mathrm{1-100~keV}\sim10^{39-44}~\mathrm{erg~s}^{-1}$ \citep{Marin2023}. Other studies have identified LLAGN with $L_\mathrm{2-10~keV}\sim10^{37-42}~\mathrm{erg~s}^{-1}$ \citep{She2018,Diaz2020}. Additionally, X-ray coronal destruction events can drastically reduce X-ray luminosity, as seen in a changing-look AGN \citep{Ricci2020}.

However, lowering the luminosity threshold may misidentify non-AGN sources such as super-Eddington accretors, which are essentially stellar mass black holes or neutron stars and are a class of high-mass X-ray binaries (HMXBs). In low-metallicity dwarf galaxies, HMXBs contribute significantly, with luminosities reaching $10^{41}\mathrm{ergs}^{-1}$ \citep{Lehmer2021,Lehmer2022}. However, these HMXBs could be either super-Eddington accretors or IMBH/LLAGN, with their 2-10 keV luminosities often exceeding $10^{39}\mathrm{ergs}^{-1}$ \citep{Swartz2011,Kaaret2017}, which are often termed ultraluminous X-ray sources (ULXs). Despite being long believed to host IMBHs, there has been a plethora of evidence showing ULXs may actually contain super-Eddington neutron star accretors \citep{Sutton2013, Bachetti2014, Karino2016, Israel2017, Pintore2018}. Therefore, relying solely on X-ray observations cannot definitively determine the presence of LLAGN, necessitating additional constraints.

Optical emission lines such as H$\alpha$, and He~II$\lambda4686$, combined with X-ray observations could be an incredibly powerful tool to constrain the origin of X-ray emission \citep{Fabrika2015,Lin2018}. 
For example, \citet{Fabrika2015} found that the line-widths of He~II$\lambda4686$ and H$\alpha$ shows almost a one-to-one correlation for super-Eddington accretors, where such high accretion rate almost always launches a very strong wind \citep{Shakura1973}.
In this paper, we adopt this approach to investigate whether green peas truly host low luminosity AGN.  

This paper is organized as follows: In Section~\ref{sec:data}, we first briefly describe the data. We then investigate the multi-wavelength data and outline the main results in Section~\ref{sec:results}, while in Section~\ref{sec:discussion} we combine the results from the observations and attempt to explain the origin of any observed multi-wavelength emission mechanisms and their potential connection to AGN. Finally in Section~\ref{sec:conclusion} we present our conclusions.

\begin{deluxetable*}{cccccccc}
\tablecaption{Host galaxy properties of green peas\label{tab:host_tab}}
\tablehead{
    \colhead{Object} & 
    \colhead{RA} & 
    \colhead{DEC} & 
    \colhead{Redshift} & 
    \colhead{SFR ($M_\mathrm{\odot}~\mathrm{yr}^{-1}$)} & 
    \colhead{12+log(O/H)} & 
    \colhead{X-ray detection} & 
    \colhead{He~II detection}
}
\startdata
J173501.2+570308 & 263.76 & 57.05 & 0.091 & 14.04 & 8.26 & True & True \\
J002101.0+005248 & 5.25 & 0.88 & 0.077 & 8.00 & 8.22 & True & True \\
J084414.2+022620 & 131.06 & 2.44 & 0.066 & 5.06 & 8.10 & True & True \\
J093813.5+542824 & 144.56 & 54.47 & 0.047 & 10.19 & 8.07 & True & True \\
J080619.5+194927 & 121.58 & 19.82 & 0.102 & 13.56 & 8.27 & True & True \\
J141059.2+430247 & 212.75 & 43.05 & 0.07 & 9.84 & 8.29 & True & True \\
J140956.6+545648 & 212.49 & 54.95 & 0.098 & 13.71 & 8.23 & True & True \\
J162410.10-002202.5 & 246.04 & -0.37 & 0.031 & 3.98 & 8.24 & True & True \\
SHOC 486 & 22.02 & -1.18 & 0.027 & 2.77 & 8.07 & True & True \\
J121903.98+152608.5 & 184.77 & 15.44 & 0.196 & 12.96 & 7.87 & False & True \\
J101629.88+073404.9 & 154.12 & 7.57 & 0.183 & 16.94 & 8.18 & False & False \\
J164235.52+422349.7 & 250.65 & 42.40 & 0.151 & 16.48 & 8.24 & False & False \\
J104645.75+302330.9 & 161.69 & 30.39 & 0.127 & 8.58 & 8.32 & False & False \\
J121932.22+213324.9 & 184.88 & 21.56 & 0.141 & 19.06 & 8.21 & False & True \\
J014721.7-091646 & 26.84 & -9.28 & 0.136 & 4.99 & 8.19 & False & False \\
J144231.37-020952.0 & 220.63 & -2.16 & 0.294 & 21.23 & 8.00 & False & False \\
J150934.17+373146.1 & 227.39 & 37.53 & 0.033 & 1.77 & 7.85 & False & True \\
J160627.53+135547.8 & 241.61 & 13.93 & 0.107 & 9.28 & 8.20 & False & True \\
J101526.38+305451.9 & 153.86 & 30.91 & 0.092 & 6.09 & 8.26 & False & False \\
J031023.94-083432.8 & 47.6 & -8.58 & 0.052 & 0.96 & 8.26 & False & False \\
J092532.36+140313.1 & 141.38 & 14.05 & 0.301 & 23.8 & 8.06 & False & False \\
J235604.67-85423.4 & 359.02 & -8.91 & 0.169 & 9.06 & 8.26 & False & False \\
J094244.23+411019.3 & 145.68 & 41.17 & 0.046 & 2.56 & 7.94 & False & False \\
J132916.55+170020.9 & 202.32 & 17.01 & 0.094 & 9.95 & 8.22 & False & True \\
J155027.78+192058.7 & 237.62 & 19.35 & 0.212 & 20.76 & 8.27 & False & False \\
J154748.99+22303.2 & 236.95 & 22.05 & 0.031 & 0.68 & 8.05 & False & False \\
J140018.93+010453.8 & 210.08 & 1.08 & 0.121 & 11.78 & 8.08 & False & False \\
J154543.55+085801.3 & 236.43 & 8.97 & 0.038 & 4.43 & 7.76 & False & True \\
\enddata
\end{deluxetable*}

\section{Observation and Data}\label{sec:data}

\subsection{Sample Selection:}
We began with the parent sample of green pea galaxies from \citet{Jiang2019}.  This consists of 1004 objects selected from Data Release 13 of the Sloan Digital Sky Survey (SDSS DR13).  The primary selection was for high equivalent width emission lines, with either the equivalent width of [O III] 5007$\mathrm{\AA}$, EW$_{5007} > 300\mathrm{\AA}$, or the equivalent width of the Balmer H$\beta$ line EW$_\mathrm{H\beta}>100\mathrm{\AA}$.  Further details of the selection are given in \citet{Jiang2019}.  
We cross matched this list with the \textit{Chandra Observations Catalog} and the \textit{Chandra Source Catalog, Version 2.0} to identify green pea galaxies where we can study X-ray emission properties.  We identified 29 green peas with \textit{CXO} observations, including 7 that are detected in the Chandra Source Catalog. 2 additional sources were detected with X-ray emission over $>3\sigma$ confidence among the remaining 22 sources.

\subsection{Optical Data:}
\hide{
We primarily assembled this sample from the spectroscopic data provided in the 7th data release of the Sloan Digital Sky Survey (SDSS DR7) \citep{Abazajian2009}. Subsequently, we crossmatched the SDSS data with the dataset presented by \citep{Cardamone2009}, resulting in 1004 low-metallicity (12+log(O/H)$\sim$8.5) dwarf galaxies as listed in the catalogue \citep{Jiang2019}. 
\textcolor{purple}{JER: Did we start from what I call ``Huan's catalog''?  If so, that does not come from cross matching Cardamone et al with Abazajian et al.   Similar but not precise or correct.}
\textcolor{purple}{JER: We need to insert here an explanation of how we go from 1004 sources to 29.  Basically I'm asking for some clarification of information in section 2.1, and reorganization of information in 2.1 and 2.2.}
\textcolor{red}{JER: We began with a list of 1004 strong emission line objects selected from SDSS Data Release N on the basis of strong H$\beta$ (EW $> x$\AA) or OIII (EW $> y$\AA) emission.  We cross matched these with X-ray catalogs, identifying 28 sources with useful depth X-ray observations.}
}

We used {\it Sloan Digital Sky Survey} optical spectra to measure 
the flux and flux error of the He~II$\lambda4686$ line for the 29 sources in our sample. We considered He~II$\lambda4686$ as detected only when the signal-to-noise ratio, $S/N_\mathrm{He~II}>3$. Among the 9 X-ray-detected sources, 8 exhibited He~II$\lambda4686$$\lambda4686$ detection over 3$\sigma$ confidence. Similarly, among the 20 X-ray non-detected sources, 9 displayed He~II$\lambda4686$  detections. One source hovered at a borderline $S/N_\mathrm{He~II}\sim2.99$. However, adopting a conservative stance, we opted to disregard this particular source due to its $S/N< 3$. 
Other emission lines used in our analysis (e.g., H$\alpha$ and [N~II]; see section~\ref{sec:Ha emission}) are similarly measured from the {\it SDSS} spectra.

\subsection{X-ray Data:}
\hide{
To search for X-ray emitting sources, we checked archival data from the \textit{Chandra Observations Catalog} and the  \textit{Chandra Source Catalog, Version 2.0}. Crossmatching our sample of dwarf galaxies with both catalogs, we found that 29 objects were observed with Chandra, in which 7 sources have X-ray detections.} 
For X-ray measurements, we used measurements from the \textit{Chandra Source Catalog, Version 2.0} for the seven sample objects contained in that catalog.  
We downloaded the data for the remaining 22 sources from the Chandra X-ray Center (CXC), and then reduced and analyzed using the Chandra Interactive Analysis of Observations (CIAO; v4.16) tools \citep{Fruscione2006}.  Firstly, we utilized the CIAO task \textit{repro} to reprocess the level-1 ACIS event file data. Then we used the task \textit{srcflux} to estimate the source counts and fluxes in the broad band, 0.5 – 7 keV and in the hard band, 2-7 keV.  For each source, we defined a circular source region centered at the source position on the broad band images, with a radius $R_\mathrm{source}$ enclosing 90\% of the PSF at 1.0 keV. Backgrounds are extracted from an annulus with $R_\mathrm{source} < R < 5 R_\mathrm{source}$, after confirming the absence of nearby bright X-ray sources. 
Out of the 22 observations, J162410.10-002202.5  and SHOC 486 are detected with significant X-ray emission (over $3\sigma$ confidence) in both broad and hard bands. Therefore, for the sources with no X-ray detection, extracting background from a region between $R_\mathrm{source}-5R_\mathrm{source}$ should not introduce any uncertainty. For 162410.10-002202.5 and SHOC 486, the background may be slightly overestimated due. However, even with the overestimated background, a source detection over $3\sigma$ strongly suggests that the detection is indeed robust.

\begin{figure}[h]
\includegraphics[width=0.5\textwidth]{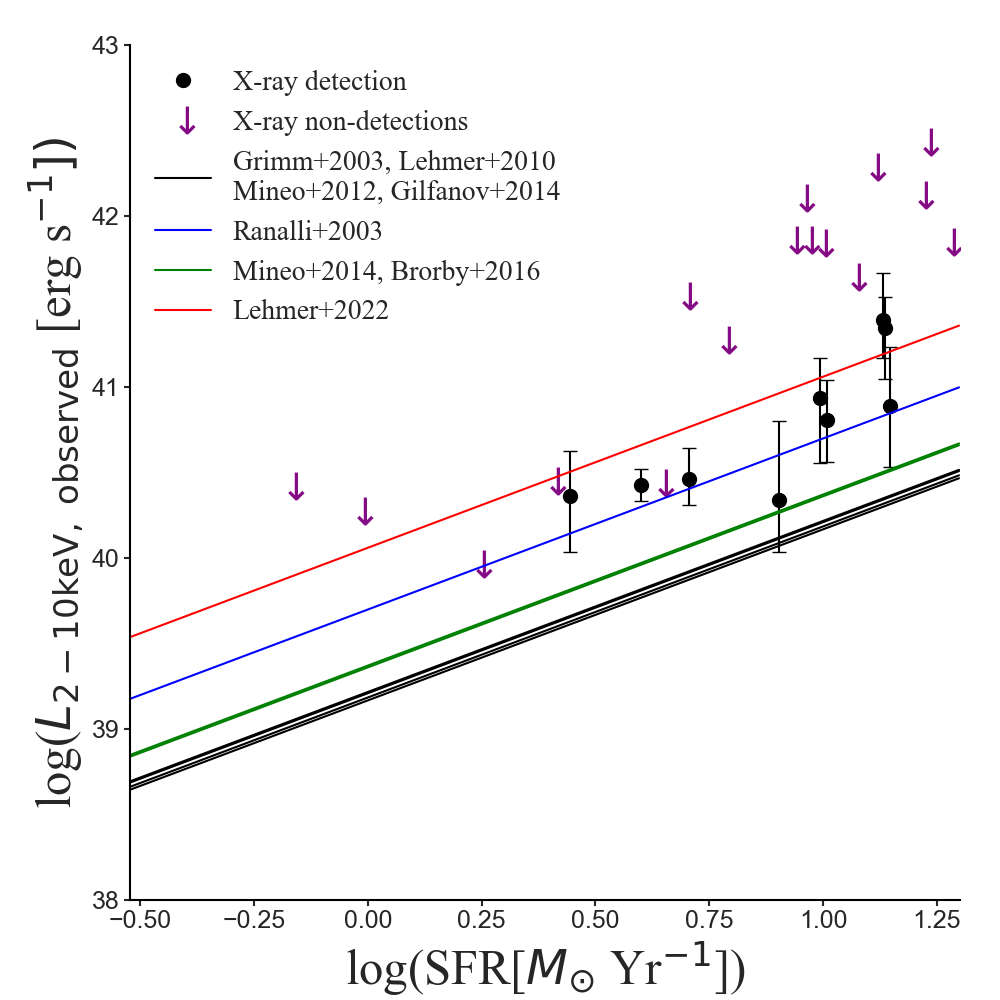}
  \caption{2-10 keV observed X-ray luminosity for our sources ($L_\mathrm{2-10~keV,~observed}$) vs. star formation rate (SFR, derived from narrow H$\alpha$ flux).  Curves show the expected relation of $L_\mathrm{2-10~keV}$ and SFR from the literature for several classes of galaxy.  X-ray detections are shown with $1\sigma$ uncertainties, and nondetections are plotted as XX\% confidence upper limits.  If our sources are similar to LIRGs, then the observed X-ray emission cannot be described by XRBs alone. However, using the low metallicity galaxy calibration, $L_\mathrm{2-10~keV,~observed}$ almost agrees with $L_\mathrm{2-10~keV,~observed}$.}
\label{fig:fig_Xray}
\end{figure}

The low net counts restrict us from performing any efficient spectrum analysis. This makes the number of sources with X-ray detection over 3$\sigma$ confidence to be 9. The host galaxy properties of these sources are listed in Table~\ref{tab:host_tab}.
We present the X-ray flux values in the 0.5-7 keV, 2-7 keV bands for the sources with  X-ray detections and $3\sigma$ upper limits for the non-detections in Table~\ref{tab:xray}. We have listed the 2-10 keV X-ray luminosity values (or the upper limits) for these sourcs as well.





\begin{deluxetable*}{ccccc}
\tablecaption{X-ray properties of green peas\label{tab:xray}}
\tablehead{
    \colhead{Object} & 
    \colhead{$F_\mathrm{2-7~keV}$} & 
    \colhead{$F_\mathrm{0.5-7~keV}$} & 
    \colhead{log($L_\mathrm{2-10~keV}$)} & 
    \colhead{log$\beta$=log($L_\mathrm{0.5-8~keV}$/SFR)} \\
    & \colhead{[$10^{-15}~\mathrm{erg~s}^{-1}~\mathrm{cm}^{-2}$]} & 
    \colhead{[$10^{-15}~\mathrm{erg~s}^{-1}~\mathrm{cm}^{-2}$]} & 
    \colhead{[$\mathrm{erg~s}^{-1}$]} & 
    \colhead{[$\mathrm{erg~s}^{-1}~(M_\mathrm{\odot}~\mathrm{yr}^{-1})^{-1}$]}
}
\startdata
J173501.2+570308 & $2.61_{-2.07}^{+2.06}$ & $4.31_{-1.79}^{+1.79}$ & $40.89_{-0.35}^{+0.36}$ & $39.86_{-0.19}^{+0.25}$ \\
J002101.0+005248 & $1.04_{-0.86}^{+0.86}$ & $2.8_{-1.05}^{+1.05}$ & $40.34_{-0.46}^{+0.3}$ & $39.76_{-0.19}^{+0.15}$ \\
J084414.2+022620 & $1.96_{-0.77}^{+0.77}$ & $2.98_{-1.18}^{+1.18}$ & $40.46_{-0.18}^{+0.16}$ & $39.84_{-0.17}^{+0.2}$ \\
J093813.5+542824 & $8.52_{-3.93}^{+3.93}$ & $13.5_{-3.7}^{+3.49}$ & $40.81_{-0.24}^{+0.25}$ & $39.89_{-0.11}^{+0.15}$ \\
J080619.5+194927 & $6.53_{-2.96}^{+2.8}$ & $9.72_{-2.64}^{+2.64}$ & $41.39_{-0.27}^{+0.23}$ & $40.33_{-0.15}^{+0.12}$ \\
J141059.2+430247 & $5.05_{-2.46}^{+2.46}$ & $9.75_{-2.04}^{+2.05}$ & $40.93_{-0.24}^{+0.38}$ & $40.12_{-0.09}^{+0.09}$ \\
J140956.6+545648 & $6.35_{-2.54}^{+3.54}$ & $11.7_{-2.43}^{+2.27}$ & $41.35_{-0.18}^{+0.3}$ & $40.37_{-0.1}^{+0.09}$ \\
J162410.10-002202.5 & $8.32_{-1.8}^{+1.9}$ & $20.2_{-2.2}^{+1.9}$ & $40.43_{-0.09}^{+0.1}$ & $40.11_{-0.05}^{+0.04}$ \\
SHOC 486 & $9.4_{-4.6}^{+5.91}$ & $14.3_{-7.0}^{+9.1}$ & $40.36_{-0.26}^{+0.33}$ & $40.0_{-0.26}^{+0.22}$ \\
J121903.98+152608.5 & $<12.42$ & $<18.93$ & $<42.29$ & $<41.25$ \\
J101629.88+073404.9 & $<20.04$ & $<30.54$ & $<42.43$ & $<41.28$ \\
J164235.52+422349.7 & $<14.96$ & $<22.8$ & $<42.12$ & $<40.98$ \\
J104645.75+302330.9 & $<11.95$ & $<18.21$ & $<41.86$ & $<41.0$ \\
J121932.22+213324.9 & $<9.24$ & $<14.08$ & $<41.85$ & $<40.64$ \\
J014721.7-091646 & $<4.86$ & $<7.4$ & $<41.53$ & $<40.91$ \\
J144231.37-020952.0 & $<11.95$ & $<18.21$ & $<42.67$ & $<41.42$ \\
J150934.17+373146.1 & $<2.63$ & $<4.0$ & $<39.96$ & $<39.79$ \\
J160627.53+135547.8 & $<17.13$ & $<26.1$ & $<41.86$ & $<40.97$ \\
J101526.38+305451.9 & $<6.22$ & $<9.48$ & $<41.27$ & $<40.57$ \\
J031023.94-083432.8 & $<2.08$ & $<3.17$ & $<40.27$ & $<40.37$ \\
J092532.36+140313.1 & $<6.13$ & $<9.34$ & $<42.41$ & $<41.11$ \\
J235604.67-85423.4 & $<11.29$ & $<17.21$ & $<42.1$ & $<41.22$ \\
J094244.23+411019.3 & $<3.97$ & $<6.05$ & $<40.45$ & $<40.12$ \\
J132916.55+170020.9 & $<21.83$ & $<33.27$ & $<41.84$ & $<40.92$ \\
J155027.78+192058.7 & $<13.74$ & $<20.94$ & $<42.41$ & $<41.17$ \\
J154748.99+22303.2 & $<8.14$ & $<12.4$ & $<40.42$ & $<40.67$ \\
J140018.93+010453.8 & $<8.05$ & $<12.27$ & $<41.64$ & $<40.65$ \\
J154543.55+085801.3 & $<5.82$ & $<8.88$ & $<40.44$ & $<39.87$ \\
\enddata
\end{deluxetable*}

\section{Analysis and results}\label{sec:results}

\subsection{X-ray emission:}\label{sec:X-ray}
We employ the Chandra-based Portable, Interactive Multi-Mission Simulator (PIMMS) to convert the 0.2-5 keV X-ray fluxes to 2-10 keV unabsorbed fluxes by applying a correction factor of approximately 1.944. This correction is based on a photon index of $\Gamma$ = 1.47, a typical value for starburst galaxies \citep{Rephaeli1995}, and a galactic hydrogen column density $N_\mathrm{H}=3\times10^{20}~\mathrm{cm}^{-2}$. The estimated 2-10 keV X-ray luminosity for these sources spans a range of $L_\mathrm{2-10~keV,~observed} = 7.8\times10^{40}-1.07\times10^{41}~\mathrm{erg~s}^{-1}$. These $L_\mathrm{2-10~keV}$ values are at least an order of magnitude higher than those typically observed in dwarf galaxies hosting AGN \citep{Reines2014,Mezcua2018,Latimer2021}, where $L_\mathrm{2-10~keV,~observed}<10^{40}~\mathrm{erg~s}^{-1}$. However, the SFR values in our sources are also an order of magnitude higher than their sources. In starburst galaxies, it is well-established that the X-ray binary population contribute to the observed X-ray emission significantly and that it increases with the galactic SFR. Therefore, it is necessary to estimate the 2-10 keV luminosity attributable to X-ray binaries $L_\mathrm{2-10~keV,~XRB}$.

Despite both low mass (LMXBs) and high mass X-ray binaries (HMXBs) contributing to 2-10 keV X-ray emission, the LMXB contribution is negligible compared to the HMXBs as documented by several studies \citep{Grimm2003,Mineo2014,Svoboda2019,Lehmer2022}. We conduct a comparative analysis of the 2-10 keV X-ray luminosity of our sources with the X-ray binary (XRB) contribution, employing various $L_{2-10~keV}$-SFR-Z calibrations sourced from the existing literature.

\citet{Ranalli2003} established a correlation between the star formation rates (SFRs) of a selected group of nearby star-forming galaxies and their total X-ray luminosity in the 2-10 keV range, denoted as $L_\mathrm{2-10~keV,~observed} \sim 5\times10^{39}~\mathrm{SFR}~(M_{\odot}~\mathrm{Yr}^{-1})$. It is crucial to acknowledge that in this context, SFR values are derived only from stars with masses exceeding 5$M_{\odot}$.
A study by \citet{Grimm2003} suggested an XRB luminosity scaling with the SFR, expressed as $L_\mathrm{2-10~keV,~observed}\sim  1.65\times10^{39}~\mathrm{SFR}~(M_{\odot}~\mathrm{Yr}^{-1})$ in the 2-10 keV band. 
\citet{Lehmer2010} prescribed a calibration very similar to \citet{Grimm2003}, $L_\mathrm{2-10~keV,~observed}\sim  1.62\times10^{39}~\mathrm{SFR}~(M_{\odot}~\mathrm{Yr}^{-1})$ for luminous infrared galaxies (LIRGs). In the mid-infrared color-color diagram, the sources in our sample occupy similar regions as the luminous infrared galaxies (LIRGs) or ultra-luminous infrared galaxies (ULIRGs) as shown by \citet{Harish2021}.

\begin{deluxetable*}{cccccc}
\tablecaption{Emission line galaxy properties of green peas}\label{tab:eline_tab}
\tablehead{
    \colhead{Object} & 
    \colhead{$\sigma_\mathrm{H\alpha,~broad}$} & 
    \colhead{$\sigma_\mathrm{He~II\lambda4686}$} & 
    \colhead{$L_\mathrm{H\alpha,~broad}$} & 
    \colhead{$L_\mathrm{He~II\lambda4686}$} & 
    \colhead{$S/N(\mathrm{He~II\lambda4686}$)}\\
    & \colhead{[$\mathrm{km~s}^{-1}$]} & 
    \colhead{[$\mathrm{km~s}^{-1}$]} & 
    \colhead{[$\mathrm{erg~s}^{-1}$]} & 
    \colhead{[$\mathrm{erg~s}^{-1}$]} 
}
\startdata
J173501.2+570308 & 102$\pm$19 & 448$\pm$186 & 42.07$\pm$0.27 & 39.74$\pm$0.3 & 3.35 \\
J002101.0+005248 & 170$\pm$15 & 502$\pm$112 & 41.44$\pm$0.2 & 39.97$\pm$0.22 & 4.49 \\
J084414.2+022620 & 131$\pm$26 & 126$\pm$40 & 41.42$\pm$0.45 & 39.25$\pm$0.27 & 3.69 \\
J093813.5+542824 & 181$\pm$34 & 162$\pm$23 & 41.24$\pm$0.22 & 39.43$\pm$0.09 & 10.7 \\
J080619.5+194927 & 173$\pm$11 & 363$\pm$123 & 41.75$\pm$0.18 & 40.0$\pm$0.23 & 4.28 \\
J141059.2+430247 & 114$\pm$20 & 560$\pm$115 & 41.74$\pm$0.5 & 39.9$\pm$0.21 & 4.79 \\
J140956.6+545648 & 210$\pm$45 & 279$\pm$47 & 42.03$\pm$0.24 & 40.05$\pm$0.12 & 8.59 \\
J162410.10-002202.5 & 90$\pm$30 & 525$\pm$139 & 40.32$\pm$3.0 & 39.46$\pm$0.2 & 5.01 \\
SHOC 486 & 82$\pm$44 & 124$\pm$21 & 41.49$\pm$0.48 & 39.12$\pm$0.09 & 11.4 \\
J121903.98+152608.5 & 227$\pm$60 & 104$\pm$29 & 41.9$\pm$0.22 & 40.19$\pm$0.21 & 4.76 \\
J121932.22+213324.9 & 193$\pm$12 & 252$\pm$70 & 41.89$\pm$0.17 & 39.96$\pm$0.28 & 3.55 \\
J150934.17+373146.1 & 56$\pm$36 & 102$\pm$45 & 41.1$\pm$47.66 & 39.02$\pm$0.11 & 8.71 \\
J160627.53+135547.8 & 189$\pm$12 & 245$\pm$85 & 41.73$\pm$0.16 & 39.76$\pm$0.25 & 4.06 \\
J132916.55+170020.9 & 198$\pm$42 & 821$\pm$241 & 41.81$\pm$0.22 & 40.21$\pm$0.26 & 3.88 \\
J154543.55+085801.3 & 57$\pm$15 & 85$\pm$9 & 41.54$\pm$46.1 & 39.48$\pm$0.05 & 18.92 \\
\enddata
\end{deluxetable*}

\begin{figure}[h!]
\includegraphics[width=0.5\textwidth]{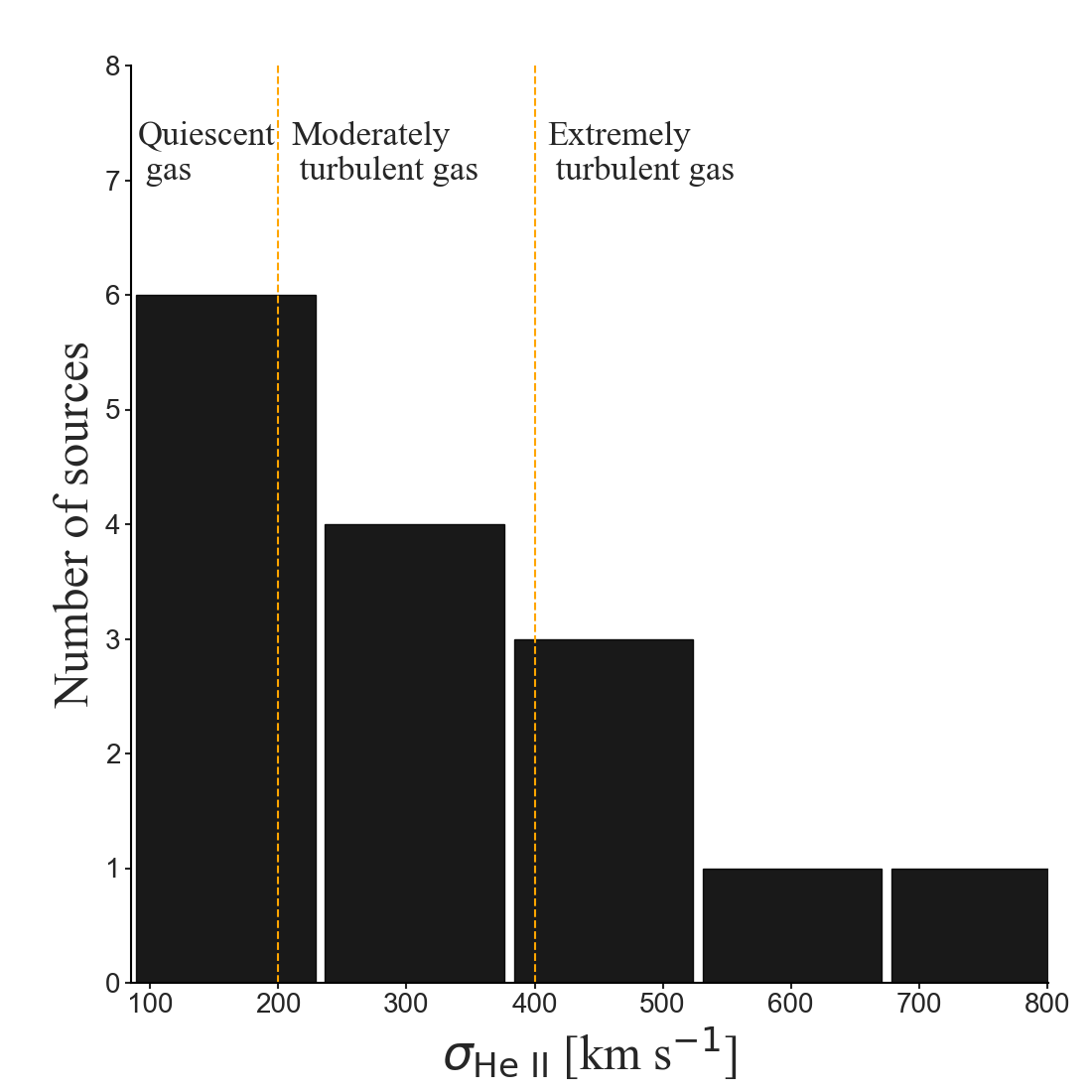}
  \caption{A histogram of the velocity dispersion ($\sigma_\mathrm{He~II\lambda4686}$) of the 15 sources where the S/N of the He~II$\lambda4685$ emission line emission line is $>3$, $\sigma_\mathrm{He~II\lambda4686}$ . The yellow vertical dashed lines are at $\sigma_\mathrm{He~II\lambda4686}=200\mathrm{km~s}^{-1}$ and $400\mathrm{km~s}^{-1}$, and represent boundaries between sources where the He~II$\lambda4686$ gas kinematics is quiescent ($\sigma_\mathrm{He~II\lambda4686}<200\mathrm{km~s}^{-1}$), moderately turbulent ($\sigma_\mathrm{He~II\lambda4686}=200-400\mathrm{km~s}^{-1}$), and highly turbulent gas ($\sigma_\mathrm{He~II\lambda4686}>400\mathrm{km~s}^{-1}$). 
   }
\label{fig:fig_HeII_hist}
\end{figure}

\citet{Mineo2012} considered the contribution from stars within the mass range of 0.1-1000$M_{\odot}$, resulting in $L_\mathrm{~2-10~keV,~XRB} \sim 1.53\times10^{39}~\mathrm{SFR}~(M_{\odot}~\mathrm{Yr}^{-1})$. Notably, their study focused solely on resolved star-forming galaxies.
\citet{Gilfanov2014} observed a closely analogous correlation, with $L_\mathrm{~2-10~keV,~XRB} \sim 1.47\times10^{39}~\mathrm{SFR}~(M_{\odot}~\mathrm{Yr}^{-1})$. Conversely, \citet{Mineo2014} studied the X-ray properties of compact (unresolved) star-forming galaxies, yielding significantly higher X-ray luminosity values compared to \citet{Mineo2012}: $L_\mathrm{~2-10~keV,~XRB}\sim2.35\times10^{39}~\mathrm{SFR}~(M_{\odot}~\mathrm{Yr}^{-1})$. It is pertinent to note that the green peas within our sample, being unresolved by Chandra, warrant a comparison of their X-ray emission not only with resolved sources but also with unresolved ones.
The effect of metallicity introduces additional complexity in understanding the X-ray emission in starburst galaxies. \citet{Brorby2016,Lehmer2019,Lehmer2022} suggested that in low-metallicity galaxies, the X-ray luminosity from HMXBs increase by $\sim$ an order of magnitude than normal galaxies. \citet{Brorby2016} suggested a calibration among $L_{2-10~keV}$, SFR and metallicity Z. Using the median  metallicity for our sources $Z\sim0.3Z_{\mathrm{\odot}}$, the \citet{Brorby2016} relation translates to  $L_\mathrm{~2-10~keV,~XRB} \sim 2.29\times10^{39}~\mathrm{SFR}~(M_{\odot}~\mathrm{Yr}^{-1})$, which is very similar to what \citet{Mineo2014} derived. 
In each of these calibrations, the X-ray luminosity calibrations from XRBs fail to produce the total observed X-ray luminosity in all of our sources as seen in Fig.~\ref{fig:fig_Xray}.   
\citet{Lehmer2022} has suggested that in low-metallicity ($7<12+\mathrm{log(O/H)}<9.2$) dwarf galaxies, the population of ultraluminous X-ray sources or ULX ($L_\mathrm{2-10~keV}>2\times10^{39}~\mathrm{erg~s}^{-1}$)- a subset of HMXB rises significantly and they dominantly contribute to the observed X-ray emission. The authors noted that these ULXs may contain intermediate mass black holes, black hole or neutron star binaries. 

\begin{figure}[h]
\includegraphics[width=0.5\textwidth]{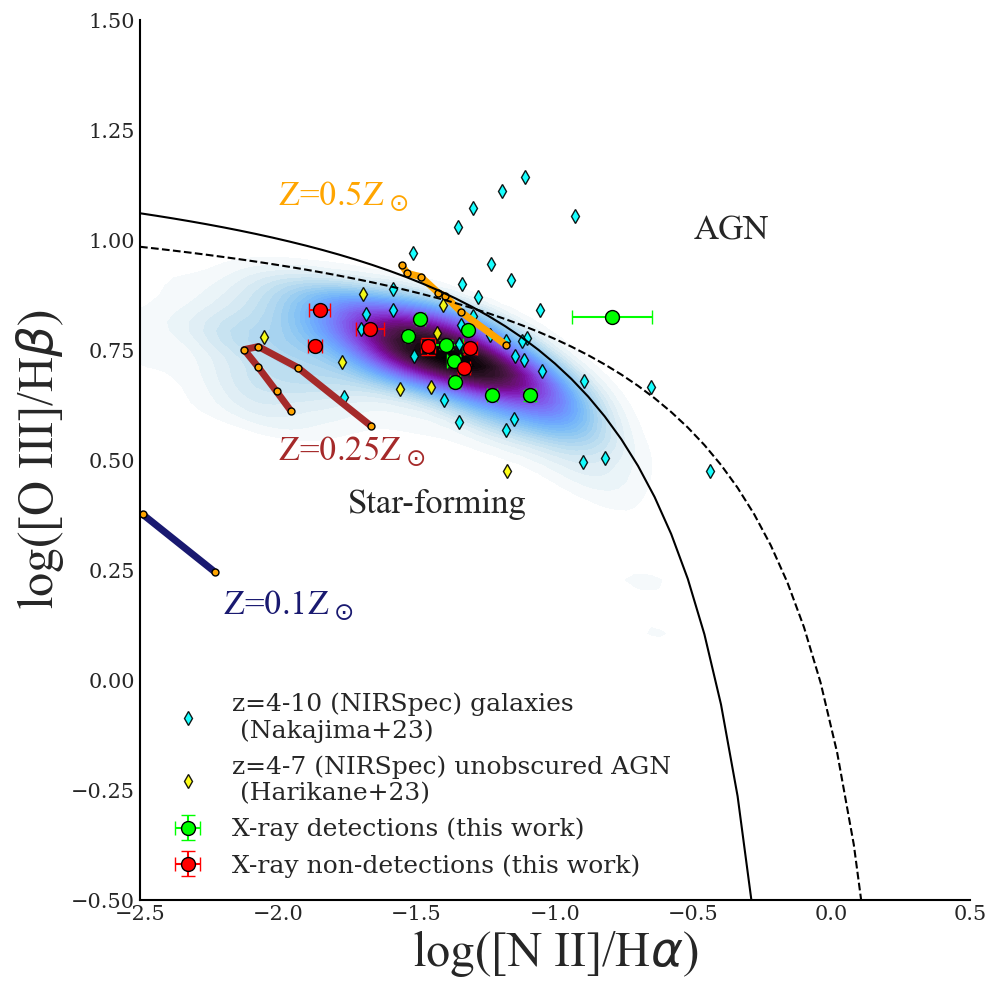}
  \caption{BPT diagram showing the emission line ratios of [O~III]/H$\beta$ plotted against those of [N~II]/H$\alpha$ for the parent sample of green peas \citep{Yang2016}, along with the 15 green peas with $>3\sigma$ He~II detection. The green circles denote the sources where X-ray emission has been detected over $>3\sigma$ confidence and the red circles represent the ones where X-ray emission has not been detected. The black solid and dashed lines represent the boundary between star-forming galaxies and AGN as per \citet{Kewley2001,Kauffmann2003}. Low metallicity AGN narrow-line regions are depicted according to \citet{Groves2006}, with different solid lines representing models for AGN host galaxies with metallicities of Z = 0.5$Z_\mathrm{\odot}$ (orange line), 0.25$Z_\mathrm{\odot}$ (brown line), and 0.1$Z_\mathrm{\odot}$ (blue line). The orange data points on these lines represent models with different intensity and ionization parameters (see Table 2 of \citet{Groves2006}). Most of our sources lie between the models for Z = 0.25$Z_\mathrm{\odot}$ and Z = 0.5$Z_\mathrm{\odot}$.
   }
\label{fig:fig_BPT}
\end{figure}

Adopting the calibration for 12 + log(O/H) = 8.2, which corresponds to the average metallicity for our sources, we do notice that 2-10 keV X-ray luminosity can be described by the calibration of \citet{Lehmer2022}. However, in J0801619.5$+$194927 and J140956.6$+$54568, $L_\mathrm{2-10~keV}$ exceeds $10^{41}~\mathrm{erg~s}^{-1}$. Sources with $L_\mathrm{2-10~keV}>10^{41}~\mathrm{erg~s}^{-1}$ are often believed to host intermediate mass black holes (IMBH, $M_\mathrm{BH}>100~M_\mathrm{\odot}$)\citep{Barrows2019}. Therefore, these two out of 9 sources are strong candidates to host IMBH.

\subsection{Optical emission lines:}\label{sec:optical emission}

\subsubsection{He~II$\lambda4686$ emission:}\label{sec:HeII emission}

For the 15 sources where $S/N_\mathrm{He~II}>3$, we fitted the emission line spectra from 4640-4740$\mathrm{\AA}$, encompassing the Fe~III$\lambda4658$, He~II$\lambda4686$ , and Ar~IV$\lambda4711$ emission lines. Modelling multiple emission lines together yielded a more robust fit. We coupled the emission line centers of the emission lines and the line-widths for Fe~III$\lambda4658$ and Ar ~IV$\lambda4711$. We noticed that the He~II$\lambda4686$ lines are broader than Fe~III$\lambda4658$ and Ar ~IV$\lambda4711$. Attempting to couple the line-widths and line-centers of Fe~III$\lambda4658$, He~II$\lambda4686$ , and Ar~IV$\lambda4711$, and adding a second broad Gaussian to He~II$\lambda4686$ , did not converge the fit. The fit only converged when we did not couple the line-width of He~II$\lambda4686$ to the other two lines.

We report that the He~II$\lambda4686$ emission line exhibits a great diversity in their line-width, with 1/3 of the total sample shows highly perturbed gas with$\sigma_\mathrm{He~II}>400\mathrm{km~s}^{-1}$. On the contrast, narrow, nebular He~II$\lambda4686$ emission is clearly visible with $\sigma_\mathrm{He~II}<200\mathrm{km~s}^{-1}$ as seen in Fig.~\ref{fig:fig_HeII_hist}. 

\subsubsection{H$\alpha$ emission}\label{sec:Ha emission}
To understand the astrophysical origin of the X-ray emission, it is necessary to verify if the H$\alpha$ and He~II$\lambda4686$ shows a one-to-one relation, as that would directly show a presence of an super-Eddington accretor.
We fit the 6530-6600$\mathrm{\AA}$ rest-frame wavelength covering H$\alpha$+[N~II]$\lambda\lambda6548,6584$ with a multi-Gaussian model where the narrow Gaussian describes the nuclear star-cluster emission and the broad Gaussian may be due to supernovae-driven winds. The line-centers and line-widths for all H$\alpha$ and [N~II] lines were coupled to each other  for both the narrow and broad Gaussian complexes. Additionally, we constrained the [N~II]$\lambda6584$/[N~II]$\lambda6548$=2.61 \citep{Singha2021,Singha2023,Winkel2023}.
All of our sample shows the broad H$\alpha$ line to be moderately turbulent at most with $\sigma_\mathrm{H\alpha}<300~\mathrm{km~s}^{-1}$. These values are in agreement with the work of \citet{Wood2015}, where the authors found that, $\sigma_\mathrm{H\alpha}=70-300~\mathrm{km~s}^{-1}$ due to such winds. However, other possibilities such as the gravitational potential of a massive black hole or additional star cluster (in J173501.2+570308 and SHOC~486) cannot be excluded either.
The values of the velocity dispersion and the luminosity of the broad components in H$\alpha$ and He~II$\lambda4686$ are listed in Table.~\ref{tab:eline_tab}. All the optical spectra are presented in the Appendix.

\begin{figure}[h]
\includegraphics[width=0.5\textwidth]{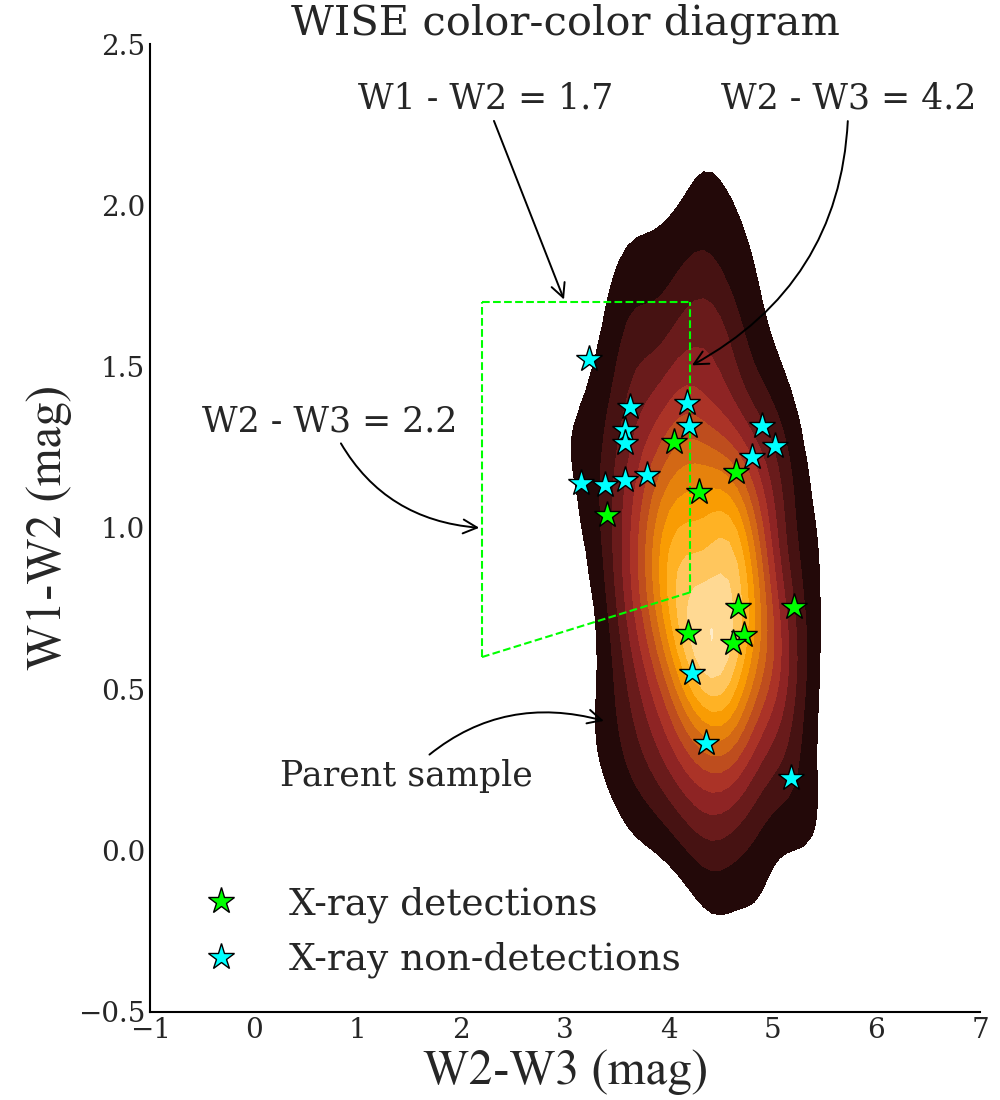}
  \caption{The WISE survey mid-infrared color-color plane used to identify candidate AGN among the GPs. The red contours describe the parent sample. The WISE color plane acts as a classification tools for different types of astrophysical objects, including stars, AGN and LIRGs etc.  \citep{Wright2010}. The dashed box is the AGN selection region as per \cite{Jarrett2011}.
   }
\label{fig:fig_MIR}
\end{figure}

\subsubsection{Ionization mechanism}\label{sec: BPT}

A popular diagnostic to understand the origin of the astrophysical processes responsible for multi-wavelength emission is the optical emission line diagnostics \citep{Baldwin1981}, using [O~III]/$H\beta$ and [N~II]/H$\alpha$ emission line ratios. To estimate the line ratio [O~III]/$H\beta$, we fit the 4840-5020$\mathrm{\AA}$ rest-frame wavelength range covering H$\beta$ and [O~III]$\lambda\lambda4959,5007$ with a single-Gaussian model and a linear model for the continuum \citep{Singha2021,Singha2022}, which was sufficient to describe the emission line features.

In these sources [O~III]/$H\beta$ and [N~II]/H$\alpha$, emission line ratios are similar to those of redshift $z > 4$ AGN detected by JWST/NIRSpec \citep{Nakajima2023,Harikane2023}, despite falling in the star-forming region of the BPT diagram (see Fig.~\ref{fig:fig_BPT}). Only SHOC 486, a source with X-ray detection, falls in the AGN-photoionization regime. \citet{Groves2006} proposed that in low-metallicity AGN, the emission line ratios could mimic those of pure star-forming galaxies. In Fig.~\ref{fig:fig_BPT}, we notice that 14 out of 15 sources with S/N(He~II) $> 3$ reside within the region described by the narrow-line region models for metallicity values of $0.25Z_\mathrm{\odot}$ and $0.5Z_\mathrm{\odot}$. Given that the median metallicity for these sources is $\sim0.32Z_\mathrm{\odot}$, their emission line ratios are consistent with the narrow-line region of low-metallicity AGN.

\subsection{Mid-infrared color:}\label{sec:MIR emission}
The Mid-infrared (MIR) color criteria \citep{Jarrett2011} using WISE observations (W$_1$ at $3.4\mu$m, W$_2$ at $4.6\mu$m, and W$_3$ at $12\mu$m) shows that the MIR colors of our sources are similar to that of ULIRGs, LIRGs and QSOs. Out of 29 sources, the MIR color is available for 25 sources including all 9 sources with $>3\sigma$ X-ray detection and 16 sources with X-ray non-detection.
Two out of nine sources with X-ray detection fall in the AGN regime as per the MIR color 
criteria suggested by \citet{Jarrett2011}, shown in Fig.~\ref{fig:fig_MIR}. Whereas for the sources with X-ray non-detection, ten out of sixteen sources lie in the AGN regime of the MIR color diagram.
\citet{Jarrett2011} suggested that a source would be consistent with AGN only if their MIR colors satisfy 2.2$<$ W2 - W3 $<$4.2, and $0.1 \times \mathrm{(W2 - W3)}+0.38 < \mathrm{W1 - W2}<1.7$. 

\begin{figure}[h]
\includegraphics[width=0.5\textwidth]{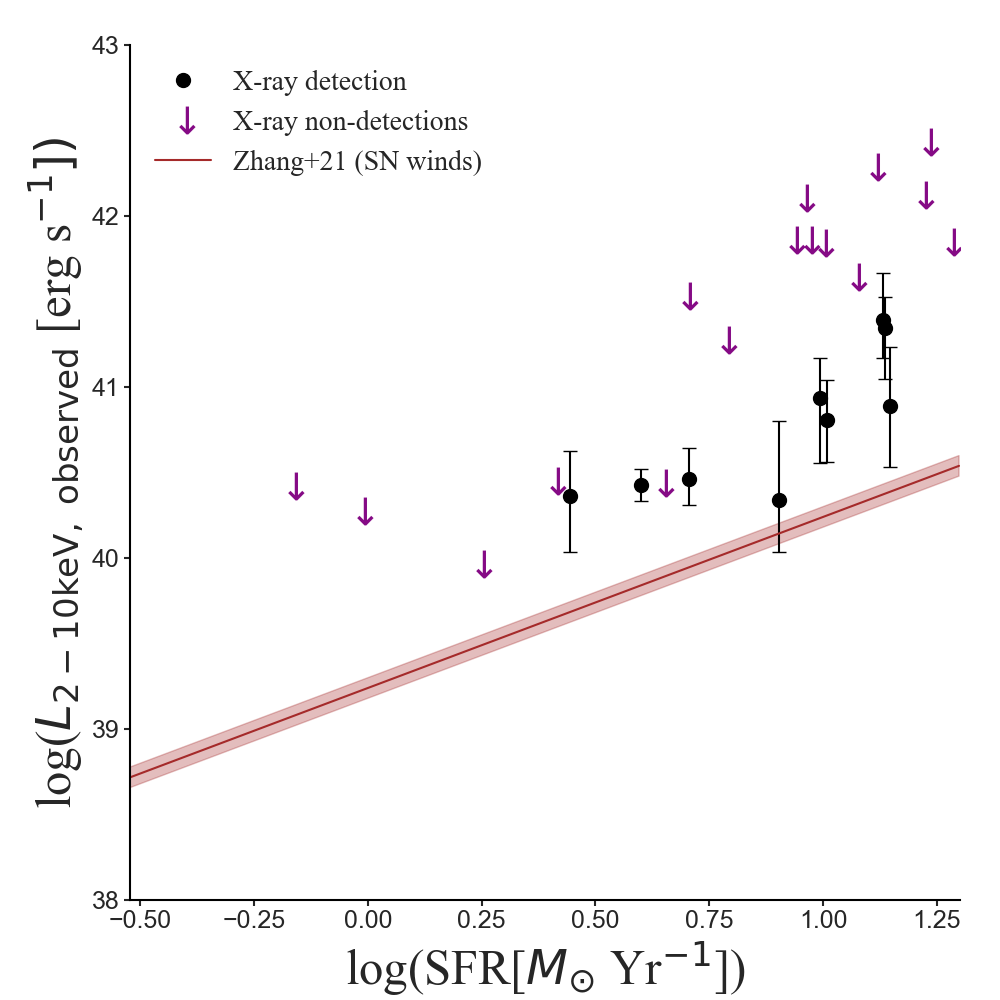}
  \caption{Similar to Fig.~\ref{fig:fig_Xray} but showing the X-ray emissions due to fast-radiative shocks from supernovae-driven winds as per \citet{Zhang2021}. The 2-10 keV X-ray luminosity of only one source falls within the range described by this calibration.  
   }
\label{fig:fig_SN_winds}
\end{figure}

In MIR wavelengths, very young compact starburst galaxies with stellar population$<$ a few Myr can result in MIR color \citep{Izotov2011,Satyapal2018}. However, 
processes driven solely by starburst activity can replicate the MIR WISE color of luminous AGNs if the ionization parameter log~U$>-0.5$ \citep{Abel2008,Satyapal2021}. Based on the optical SDSS spectra, we estimate that \textcolor{black}{$-3.1<$log~U$<-2.5$ for our sample}, with an average Log~U$_\mathrm{avg}$ of $-2.8\pm0.3$, at a 300 pc radius (assuming an electron density of 200~cm$^{-3}$), which is typically the half-light radius for these sources \citep{Kim2021}.
At such moderate Log~U values, only radiation from AGN accretion disk can produce 
these MIR colors. However, it is unclear if the MIR colors of IMBH are similar to AGN or not.
Therefore, it is entirely possible that the MIR colors are produced by Compton-thick massive black holes ($M_\mathrm{BH}>100M_\mathrm{\odot}$), whereas they almost contribute negligibly to the X-ray emission. 

\begin{figure}[h!]
\includegraphics[width=0.5\textwidth]{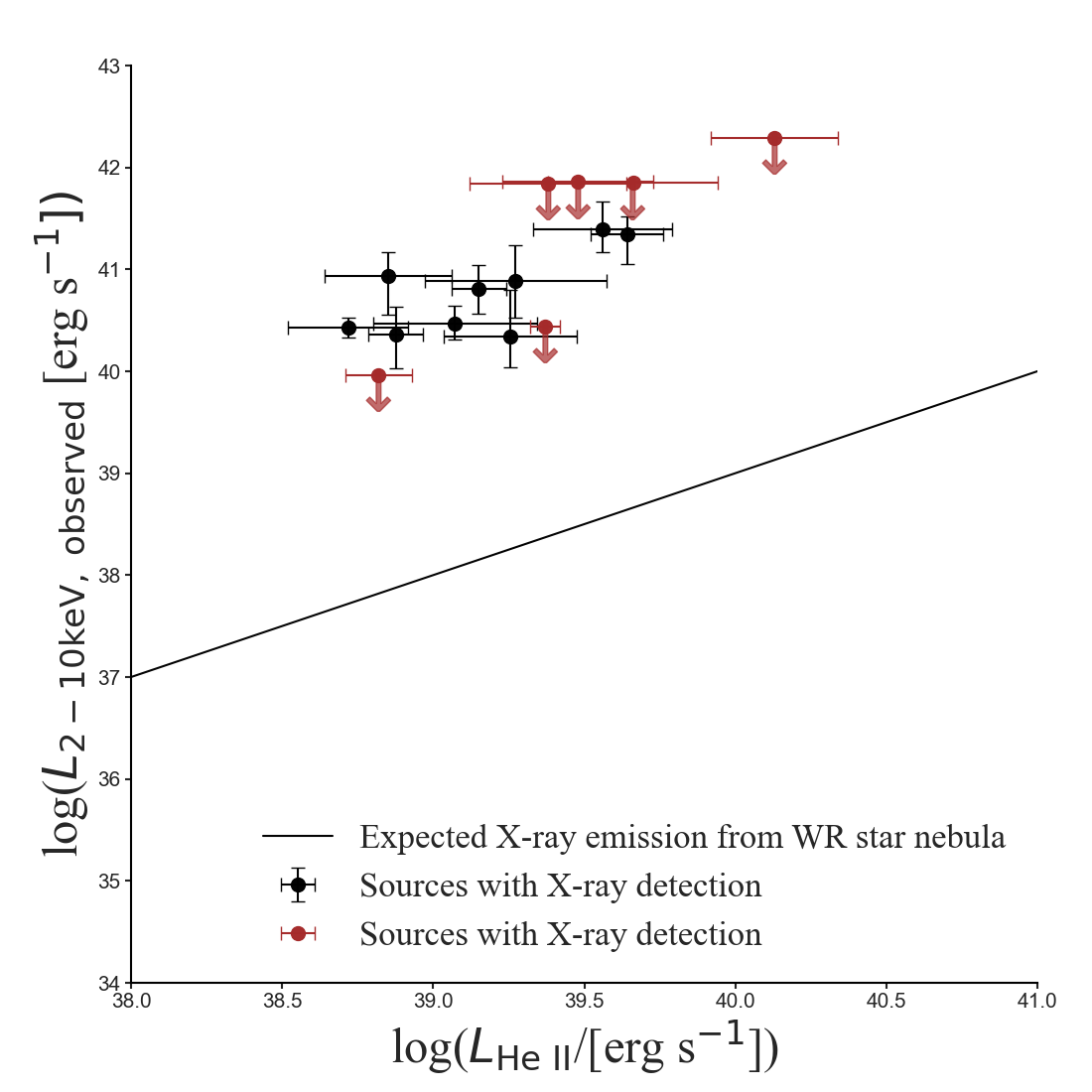}
  \caption{A plot showing He~II$\lambda4686$ vs. X-ray luminosity for the 15 sources where $S/N(\mathrm{He~II\lambda4686})>3$. The black dotted line describes the expected X-ray emission from WR stars if the observed He~II$\lambda4686$ was fully produced by them. 
   }
\label{fig:fig_HeII_WR}
\end{figure}


\section{Discussion}\label{sec:discussion}

In this section we combine the X-ray and He~II$\lambda4686$ emissions and attempt to characterize the astrophysical origin of the X-ray emission and to identify any potential signature of massive BHs.

\subsection{Stellar origin?}

\subsubsection{Fast radiative shocks:}


A popular hypothesis behind the He~II$\lambda4686$ and X-ray emission in starburst galaxies is the fast radiative shocks due to supernovae-driven winds, which manifest as broad emission line components in the Balmer lines \citep{Heckman1981,Izotov2011,Zhang2021} including H$\alpha$ \citep{Wood2015}. However, the $L_\mathrm{2-10~keV}$ produced by these shocks are about an order of magnitude lower than that seen in our sources as seen in Fig.~\ref{fig:fig_SN_winds}. The only exception is  J002101.0$+$005248 where the observed $L_\mathrm{2-10~keV}$ is described by the X-ray luminosity vs. SFR calibrations for these shocks \citep{Zhang2021}. Therefore, for the majority of our sources fast radiative shocks cannot solely describe the observed X-ray emission.

\begin{figure}[h]
\includegraphics[width=0.5\textwidth]{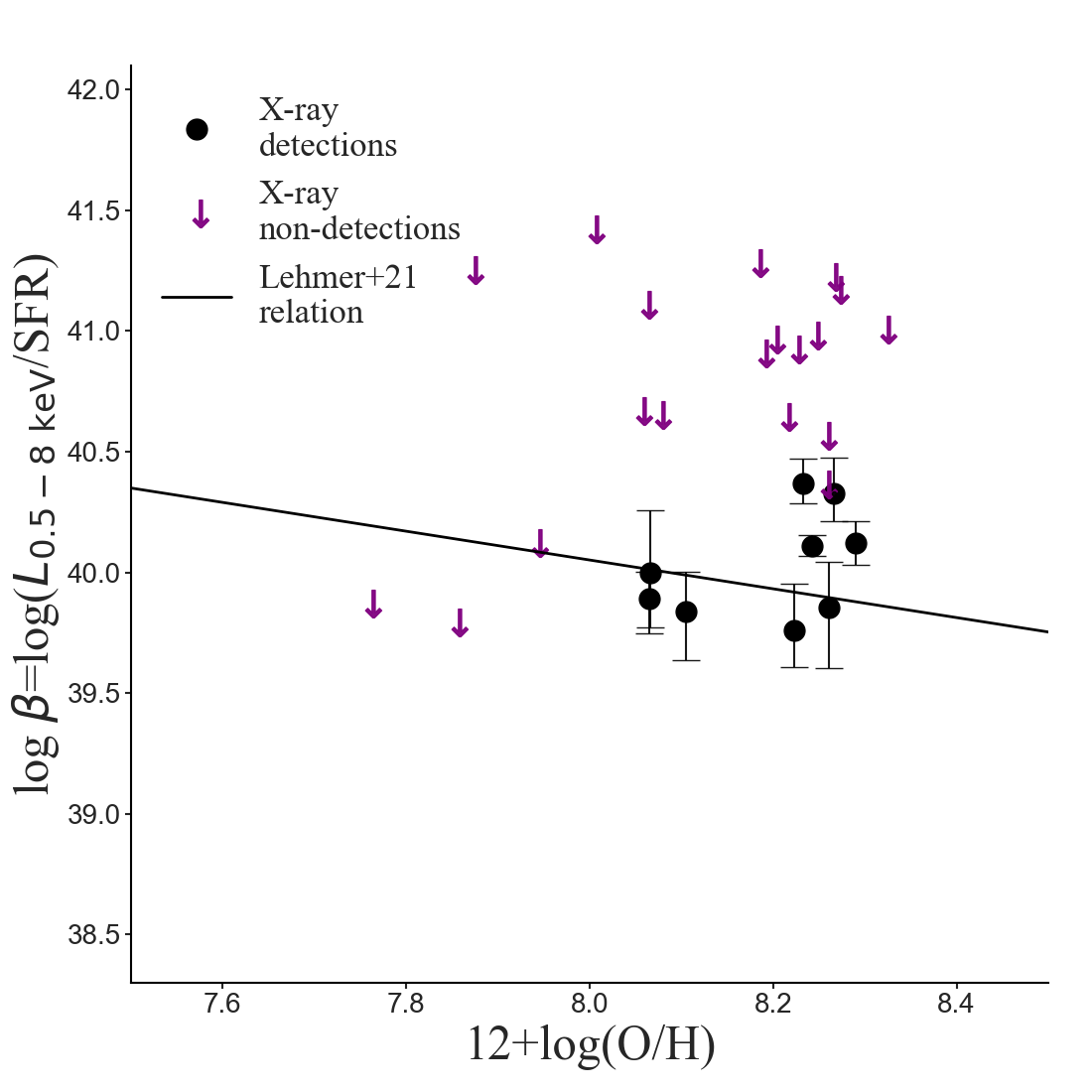}
  \caption{A plot showing $\beta$ and metallicity relation for sources with X-ray detections (in black circles) and non-detections (in purple arrows). If the X-ray emission is dominated by HMXBs, we would expect an anti-correlation between SFR and 12+log(O/H), as shown by the yellow line denoting the calibration by \citet{Lehmer2022}. However, no such anti-correlation is observed. 
   }
\label{fig:fig_SFR_Z_hist}
\end{figure}

\subsubsection{Wolf-Rayet stars:}


If we assume that both the X-ray and He~II$\lambda4686$ are produced by a Wolf-Rayet (WR) star nebula, the diversity in He~II$\lambda4686$ line profiles challenges this notion. WR stars are often characterized by their broad He~II$\lambda4686$ profile ($\sigma_\mathrm{He~II}>300~\mathrm{km~s}^{-1}$)\citep{Crowther2000,Crowther2007}. However, there are six sources where the He~II$\lambda4686$ velocity dispersion $<200~\mathrm{km~s}^{-1}$, in contrast to a broad He~II$\lambda4686$ emission profile, that one would expect from WR stars. 

The typical He~II luminosity of a WR star is $L_\mathrm{He~II}\sim10^{36}~\mathrm{erg~s}^{-1}$ \citep{Crowther2007,Crowther2023}. The observed He~II$\lambda4686$ luminosity in our sample ranges between $7\times10^{38}-3\times10^{40}~\mathrm{erg~s}^{-1}$. Assuming the entire {He~II}$\lambda4686$ nebula being produced solely by WR stars, that would give us $700-10^{4}$ WR stars in the galaxies. The typical 2-10 keV X-ray luminosity from WR stars is $10^{34}~\mathrm{erg~s}^{-1}$ \citep{White1986,Raaassen2003}. Therefore, the expected 2-10 keV X-ray luminosity from the WR star nebula would be $7\times10^{36}-10^{38}~\mathrm{erg~s}^{-1}$, which is about two orders of magnitude lower than the observed $L_\mathrm{2-10~keV}$ (see Fig.~\ref{fig:fig_HeII_WR}). Therefore, WR stars cannot produce the observed X-ray and He~II$\lambda4686$ emission.

\begin{figure}[h!]
\includegraphics[width=0.5\textwidth]{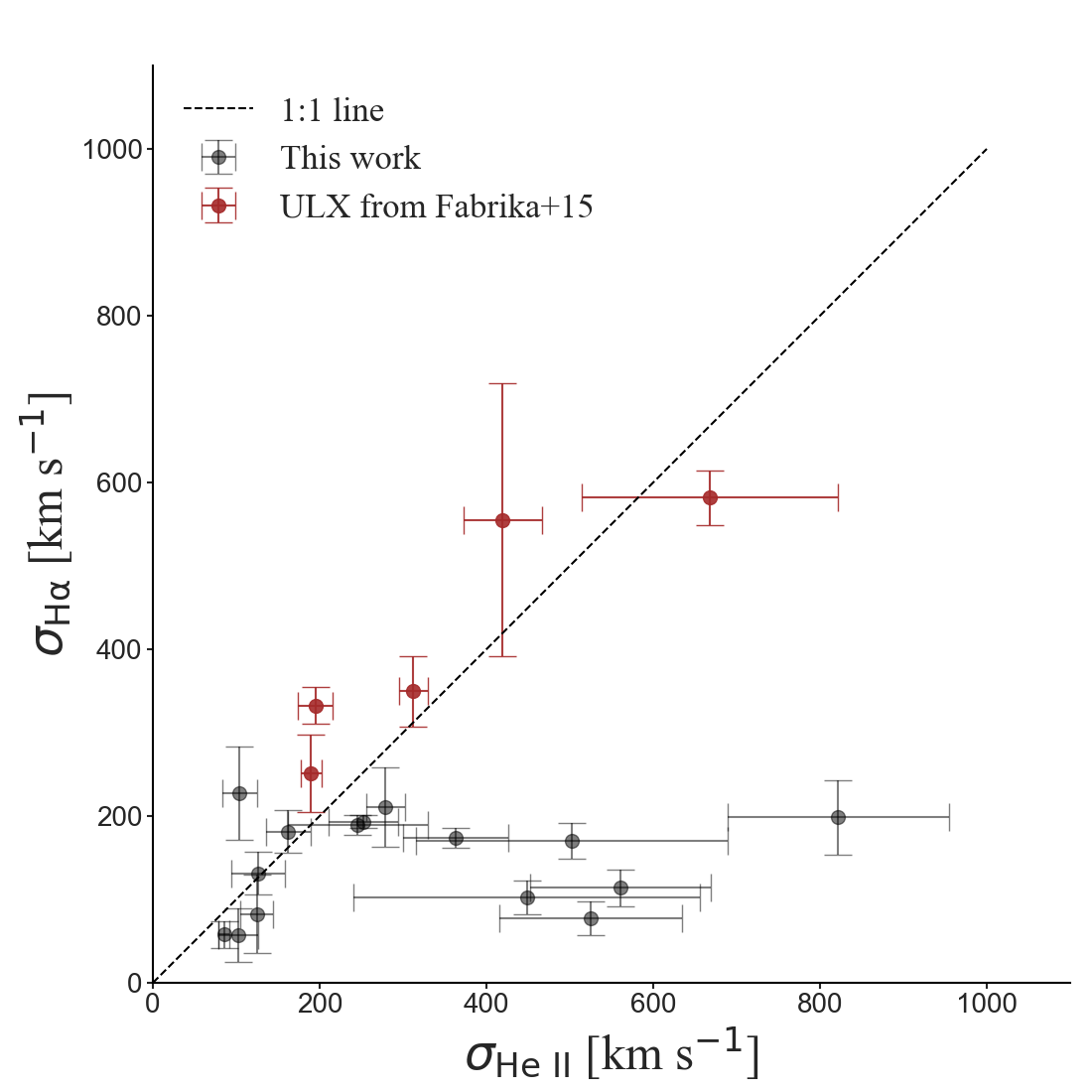}
  \caption{A scatter plot showing the velocity dispersion of the H$\alpha$ broad component ($\sigma_\mathrm{H\alpha,~broad}$) against that of He~II$\lambda4686$ ($\sigma_\mathrm{He~II\lambda4686}$). The ULX from \citet{Fabrika2015} are shown in red. The black dashed line represents where  $\sigma_\mathrm{H\alpha,~broad}=\sigma_\mathrm{He~II\lambda4686}$. In most our sources $\sigma_\mathrm{He~II\lambda4686}$ is systematically much higher than $\sigma_\mathrm{H\alpha,~broad}$, as opposed to the ULXs.
   }
\label{fig:fig_HeII_Ha_sigma}
\end{figure}

\subsection{HMXB/ULX?}


We compare the normalized 0.5-8 keV X-ray luminosity, $\beta = L_\mathrm{0.5-8~keV}$/SFR, against 12 + log(O/H) for both X-ray detected and non-detected sources, as illustrated in Fig.~\ref{fig:fig_SFR_Z_hist}. We convert the 0.5-7 keV broadband X-ray luminosity to 0.5-8 keV X-ray luminosity using a conversion factor of 1.12. We observe that among the 9 X-ray detected sources, the normalized X-ray luminosity ($\beta$) in 5 of them is consistent with an anticorrelation with 12 + log(O/H) \citep{Lehmer2021}. 
This correlation is primarily due to the large uncertainties in X-ray luminosities. \citet{Lehmer2021} suggested that in low-metallicity (12+log(O/H)=4-9.2) dwarf galaxies, the number of high-mass X-ray binaries increases with decreasing metallicity, resulting in elevated X-ray emission. The authors proposed that the majority of the X-ray emission originates from X-ray binaries with 2-10 keV X-ray luminosities exceeding $10^{38}~\mathrm{erg~s}^{-1}$, commonly known as ULXs. However, ULXs could either be (i) super-Eddington accretors or (ii) intermediate-mass black holes (IMBHs). However, in 4 other sources with $>3\sigma$ X-ray detection, there is an excess X-ray emission in comparison to the \citet{Lehmer2021} correlation.
To comprehensively understand the origin of the X-ray emission, we need to determine which of these two scenarios is most consistent with our multi-wavelength observations.


\subsubsection{Super-Eddington accretor:} To produce the observed X-ray luminosity in the 2-10 keV range ($L_\mathrm{2-10~keV}>10^{40.5}~\mathrm{erg~s}^{-1}$), both a 2 $M_\mathrm{\odot}$ neutron star (the heaviest neutron star) and a 100 $M_\mathrm{\odot}$ black hole (the heaviest possible stellar mass black hole) necessitate a super-Eddington accretion rate for sources with X-ray detection. For X-ray non-detected sources, if the mentioned upper limit of $L_\mathrm{2-10~keV}$ is their 2-10 keV X-ray luminosity, then a super-Eddington accretion is also required. In a super-Eddington accretion scenario, a strong outflow should be generated at the accretion disk surface via radiation pressure force \citep{Shakura1973,Ohsuga2005}. This wind manifests as broad He~II$\lambda4686$ and H$\alpha$ lines, with the Balmer lines expected to be broader than He~II$\lambda4686$ since they originate from regions of the wind situated farther from the source and have already gained speed \citep{Fabrika2015}. However, a study by \citet{Fabrika2015} on known super-Eddington accretors reveals that the line width of broad He~II$\lambda4686$ and H$\alpha$ would be strongly correlated with $R^{2}=0.9$. Importantly, we find a very weak correlation between the broad H$\alpha$ and He~II$\lambda4686$ line profiles with $R^{2}=0.15$ as seen in Fig.~\ref{fig:fig_HeII_Ha_sigma}. Additionally, the velocity dispersion of broad H$\alpha$ is systematically lower than that of He~II$\lambda4686$ and the broad H$\alpha$ traces the ionized gas shocked by supernova-driven winds. Therefore, it is evident that the compact X-ray emitting source is not a super-Eddington accretor. Consequently, a black hole with $M_\mathrm{BH}>100~M\mathrm{\odot}$ is necessary to produce the observed X-ray and He~II$\lambda4686$ emission. Together with X-ray, He~II emission, their mission line ratios are consistent with a low-metallicity low-luminosity AGN origin.

\subsubsection{Intermediate mass black holes/low-luminosity AGN:} By adopting a bolometric correction factor, $\kappa_\mathrm{2-10~keV}=20$, the bolometric luminosity, $L_\mathrm{bol}=\kappa_\mathrm{2-10~keV}L_\mathrm{2-10~keV}=(4\times10^{40}-7\times10^{42})~\mathrm{erg~s}^{-1}$, can be easily produced by a sub-Eddington LLAGN harboring an intermediate mass black hole (IMBH) with $M_\mathrm{BH}<10^{5}~M\mathrm{\odot}$. Assuming the broad H$\alpha$ component is due to the gravitational potential of massive BHs, the BH mass for these sources will range between $9.8\times10^{3}-4.9\times10^{5}M_\mathrm{\odot}$ \citep{Greene2005}.

\textit{Implication on reionization:} Traditionally, IMBHs have been believed to have a BH mass upper limit of $10^{5}~M_\mathrm{\odot}$ \citep{Reines2014,Barrows2019}. Recent studies suggest that IMBHs could have BH masses up to $10^{6}~M_\mathrm{\odot}$ \citep{Zuo2024}.
LLAGN can have ionizing photon production rates significantly higher than their host galaxies, each producing up to $\log(\dot{N}/s^{-1}) \sim 51$ \citep{Kubota2018, Yung2021}, and could have contributed largely to the ionizing photon budget during the Epoch of Reionization. Particularly, recent JWST detection of many high-redshift ``little red dots'' has implied a higher than expected AGN fraction, as high as 71\% \citep[e.g.][also see \citealt{Habouzit2024} for a comprehensive discussions of JWST-detected AGN in the context of state-of-the-art cosmological simulations]{Kocevski2023, Kocevski2024, Larson2023}, and has rekindled discussions on their role in cosmic IGM reionization.
While these LLAGN could be quite efficient at \textit{producing} ionizing photons, their overall contribution depends on their number density and ionizing photon escape fraction, which remains unknown. Future multi-wavelength observations including JWST, ALMA are key to our understanding on their contribution in the producing the LyC photons.

\section{Conclusion}\label{sec:conclusion}
We have searched the archival Chandra data for a sample of 900 Green Peas, with observations available only for 29 sources. To grasp the origin of X-ray emission in these sources, we examined the presence of the high-ionization emission line He~II $\lambda4686$ and analyzed the H$\alpha$ emission line profile in these 29 sources. Additionally, we supplemented the optical and X-ray observations with mid-infrared WISE observations which available for 25 out of 29 sources. We report that:
\begin{itemize}
    \item 9 out of 29 sources show X-ray detection over 3$\sigma$ confidence. The 2-10 keV X-ray luminosity for two sources exceed $10^{41}~\mathrm{erg~s}^{-1}$, making them strong candidates for IMBH.
    \item He~II$\lambda4686$ is detected over 3$\sigma$ confidence in 15 out of 29 sources. 8 out of 9 sources with X-ray detection show He~II$\lambda4686$ detection.
    \item He~II$\lambda4686$ shows a diversity in line-width, $\sigma_\mathrm{He~II\lambda4686}= \mathrm{85-821~km~s}^{-1}$. In 9 out of 15 sources $\sigma_\mathrm{He~II\lambda4686}$ exceed $\mathrm{200~km~s}^{-1}$ suggesting presence of turbulent gas.
    \item The H$\alpha$ emission line in all 15 sources with  He~II$\lambda4686$ detection, requires two-Gaussian components (including a broad Gaussian), suggesting possible presence of winds. The broad Gaussian component most likely traces the  moderately turbulent gas, $\sigma_\mathrm{H\alpha,~broad}<\mathrm{230~km~s}^{-1}$.
    \item The emission line ratios such as [O~III]/$H\beta$ and [N~II]/H$\alpha$ for these sources are consistent with the narrow-line region of low-metallicity AGN.
    \item The WISE color information is available for 25 sources, including 9 sources with X-ray detection and 16 sources with X-ray non-detection.
    2 out of 9 X-ray detected sources are consistent with AGN signature \citep{Jarrett2011}.
    10 out of 16 X-ray non-detected sources are consistent with MIR colors of AGN.
    \item The 2-10 keV X-ray luminosity produced by fast-radiative shocks from supernovae and Wolf-Rayet stars is at least an order of magnitude lower than the observed X-ray luminosity.
    \item The normalized 0.5-8 keV X-ray luminosity of 5 X-ray detected sources are consistent with HMXB signature. However, the X-ray luminosity could only be produced by either (i) a super-Eddington accretor, or (ii) an IMBH/LLAGN.
    \item The H$\alpha$ and He~II$\lambda4686$ linewidths show a weak correlation, as opposed to that expected from super-Eddington accretors.
    \item The observed X-ray could be then described by an LLAGN ($M_\mathrm{BH}=10^4-10^6M_\mathrm{\odot}$), showing that LLAGN indeed exist in these GPs.
\end{itemize}

\section*{ACKNOWLEDGEMENT}
The material is based upon work supported by NASA under award number 80GSFC21M0002.

Z.Y.Z. acknowledges the support by the National Science Foundation of China (12022303) and the China-Chile Joint Research Fund (CCJRF No. 1906). We also acknowledge the science research grants from the China Manned Space Project with NO. CMS-CSST-2021-A04, CMS-CSST-2021-A07. F.T.Y. acknowledges the support by the Funds for Key Programs of Shanghai Astronomical Obser- vatory (No. E195121009) and the Natural Science Foundation of Shanghai (Project Number: 21ZR1474300).

 Funding for the Sloan Digital Sky Survey V has been provided by the Alfred P. Sloan Foundation, the Heising-Simons Foundation, the National Science Foundation, and the Participating Institutions. SDSS acknowledges support and resources from the Center for High- Performance Computing at the University of Utah. The SDSS web site is www.sdss.org.
 
SDSS is managed by the Astrophysical Research Consortium for the Participating Institutions of the SDSS Collaboration, including the Carnegie Institution for Science, Chilean National Time Allocation Committee (CNTAC) ratified researchers, the Gotham Par- ticipation Group, Harvard University, Heidelberg University, The Johns Hopkins University, L’Ecole polytechnique fédérale de Lau- sanne (EPFL), Leibniz-Institut für Astrophysik Potsdam (AIP), Max- Planck-Institut für Astronomie (MPIA Heidelberg), Max-Planck- Institut für Extraterrestrische Physik (MPE), Nanjing University, Na- tional Astronomical Observatories of China (NAOC), New Mexico State University, The Ohio State University, Pennsylvania State University, Smithsonian Astrophysical Observatory, Space Telescope Science Institute (STScI), the Stellar Astrophysics Participation Group, Universidad Nacional Autónoma de México, University of Arizona, University of Colorado Boulder, University of Illinois at Urbana-Champaign, University of Toronto, University of Utah, Uni- versity of Virginia, Yale University, and Yunnan University.

\bibliography{bib1.bib}

\begin{thebibliography}{}
\expandafter\ifx\csname natexlab\endcsname\relax\def\natexlab#1{#1}\fi
\providecommand{\url}[1]{\href{#1}{#1}}
\providecommand{\dodoi}[1]{doi:~\href{http://doi.org/#1}{\nolinkurl{#1}}}
\providecommand{\doeprint}[1]{\href{http://ascl.net/#1}{\nolinkurl{http://ascl.net/#1}}}
\providecommand{\doarXiv}[1]{\href{https://arxiv.org/abs/#1}{\nolinkurl{https://arxiv.org/abs/#1}}}

\bibitem[{{Abel} \& {Satyapal}(2008)}]{Abel2008}
{Abel}, N.~P., \& {Satyapal}, S. 2008, \apj, 678, 686, \dodoi{10.1086/529013}

\bibitem[{{Alexander} \& {Hickox}(2012)}]{Alexander2012}
{Alexander}, D.~M., \& {Hickox}, R.~C. 2012, \nar, 56, 93,
  \dodoi{10.1016/j.newar.2011.11.003}

\bibitem[{{Atek} {et~al.}(2024){Atek}, {Labb{\'e}}, {Furtak}, {Chemerynska},
  {Fujimoto}, {Setton}, {Miller}, {Oesch}, {Bezanson}, {Price}, {Dayal},
  {Zitrin}, {Kokorev}, {Weaver}, {Brammer}, {Dokkum}, {Williams}, {Cutler},
  {Feldmann}, {Fudamoto}, {Greene}, {Leja}, {Maseda}, {Muzzin}, {Pan},
  {Papovich}, {Nelson}, {Nanayakkara}, {Stark}, {Stefanon}, {Suess}, {Wang}, \&
  {Whitaker}}]{Atek2024}
{Atek}, H., {Labb{\'e}}, I., {Furtak}, L.~J., {et~al.} 2024, \nat, 626, 975,
  \dodoi{10.1038/s41586-024-07043-6}

\bibitem[{{Bachetti} {et~al.}(2014){Bachetti}, {Harrison}, {Walton},
  {Grefenstette}, {Chakrabarty}, {F{\"u}rst}, {Barret}, {Beloborodov}, {Boggs},
  {Christensen}, {Craig}, {Fabian}, {Hailey}, {Hornschemeier}, {Kaspi},
  {Kulkarni}, {Maccarone}, {Miller}, {Rana}, {Stern}, {Tendulkar}, {Tomsick},
  {Webb}, \& {Zhang}}]{Bachetti2014}
{Bachetti}, M., {Harrison}, F.~A., {Walton}, D.~J., {et~al.} 2014, \nat, 514,
  202, \dodoi{10.1038/nature13791}

\bibitem[{{Baldwin} {et~al.}(1981){Baldwin}, {Phillips}, \&
  {Terlevich}}]{Baldwin1981}
{Baldwin}, J.~A., {Phillips}, M.~M., \& {Terlevich}, R. 1981, \pasp, 93, 5,
  \dodoi{10.1086/130766}

\bibitem[{{Barrows} {et~al.}(2019){Barrows}, {Mezcua}, \&
  {Comerford}}]{Barrows2019}
{Barrows}, R.~S., {Mezcua}, M., \& {Comerford}, J.~M. 2019, \apj, 882, 181,
  \dodoi{10.3847/1538-4357/ab338a}

\bibitem[{{Basu-Zych} \& {Scharf}(2004)}]{Basu-Zych2004}
{Basu-Zych}, A., \& {Scharf}, C. 2004, \apjl, 615, L85, \dodoi{10.1086/426390}

\bibitem[{{Birchall} {et~al.}(2022){Birchall}, {Watson}, {Aird}, \&
  {Starling}}]{Birchall2022}
{Birchall}, K.~L., {Watson}, M.~G., {Aird}, J., \& {Starling}, R.~L.~C. 2022,
  \mnras, 510, 4556, \dodoi{10.1093/mnras/stab3573}

\bibitem[{{Brorby} {et~al.}(2016){Brorby}, {Kaaret}, {Prestwich}, \&
  {Mirabel}}]{Brorby2016}
{Brorby}, M., {Kaaret}, P., {Prestwich}, A., \& {Mirabel}, I.~F. 2016, \mnras,
  457, 4081, \dodoi{10.1093/mnras/stw284}

\bibitem[{{Calhau} {et~al.}(2020){Calhau}, {Sobral}, {Santos}, {Matthee},
  {Paulino-Afonso}, {Stroe}, {Simmons}, {Barlow-Hall}, \& {Adams}}]{Calhau2020}
{Calhau}, J., {Sobral}, D., {Santos}, S., {et~al.} 2020, \mnras, 493, 3341,
  \dodoi{10.1093/mnras/staa476}

\bibitem[{{Cardamone} {et~al.}(2009){Cardamone}, {Schawinski}, {Sarzi},
  {Bamford}, {Bennert}, {Urry}, {Lintott}, {Keel}, {Parejko}, {Nichol},
  {Thomas}, {Andreescu}, {Murray}, {Raddick}, {Slosar}, {Szalay}, \&
  {Vandenberg}}]{Cardamone2009}
{Cardamone}, C., {Schawinski}, K., {Sarzi}, M., {et~al.} 2009, \mnras, 399,
  1191, \dodoi{10.1111/j.1365-2966.2009.15383.x}

\bibitem[{{Crowther}(2000)}]{Crowther2000}
{Crowther}, P.~A. 2000, \aap, 356, 191, \dodoi{10.48550/arXiv.astro-ph/0001226}

\bibitem[{{Crowther}(2007)}]{Crowther2007}
---. 2007, \araa, 45, 177, \dodoi{10.1146/annurev.astro.45.051806.110615}

\bibitem[{{Crowther} {et~al.}(2023){Crowther}, {Rate}, \&
  {Bestenlehner}}]{Crowther2023}
{Crowther}, P.~A., {Rate}, G., \& {Bestenlehner}, J.~M. 2023, \mnras, 521, 585,
  \dodoi{10.1093/mnras/stad418}

\bibitem[{{Diaz} {et~al.}(2020){Diaz}, {Ar{\'e}valo},
  {Hern{\'a}ndez-Garc{\'\i}a}, {Bassani}, {Malizia},
  {Gonz{\'a}lez-Mart{\'\i}n}, {Ricci}, {Matt}, {Stern}, {May}, {Zezas}, \&
  {Bauer}}]{Diaz2020}
{Diaz}, Y., {Ar{\'e}valo}, P., {Hern{\'a}ndez-Garc{\'\i}a}, L., {et~al.} 2020,
  \mnras, 496, 5399, \dodoi{10.1093/mnras/staa1762}

\bibitem[{{Fabrika} {et~al.}(2015){Fabrika}, {Ueda}, {Vinokurov}, {Sholukhova},
  \& {Shidatsu}}]{Fabrika2015}
{Fabrika}, S., {Ueda}, Y., {Vinokurov}, A., {Sholukhova}, O., \& {Shidatsu}, M.
  2015, Nature Physics, 11, 551, \dodoi{10.1038/nphys3348}

\bibitem[{{Fruscione} {et~al.}(2006){Fruscione}, {McDowell}, {Allen},
  {Brickhouse}, {Burke}, {Davis}, {Durham}, {Elvis}, {Galle}, {Harris},
  {Huenemoerder}, {Houck}, {Ishibashi}, {Karovska}, {Nicastro}, {Noble},
  {Nowak}, {Primini}, {Siemiginowska}, {Smith}, \& {Wise}}]{Fruscione2006}
{Fruscione}, A., {McDowell}, J.~C., {Allen}, G.~E., {et~al.} 2006, in Society
  of Photo-Optical Instrumentation Engineers (SPIE) Conference Series, Vol.
  6270, Observatory Operations: Strategies, Processes, and Systems, ed. D.~R.
  {Silva} \& R.~E. {Doxsey}, 62701V, \dodoi{10.1117/12.671760}

\bibitem[{{Giallongo} {et~al.}(2015){Giallongo}, {Grazian}, {Fiore}, {Fontana},
  {Pentericci}, {Vanzella}, {Dickinson}, {Kocevski}, {Castellano}, {Cristiani},
  {Ferguson}, {Finkelstein}, {Grogin}, {Hathi}, {Koekemoer}, {Newman}, \&
  {Salvato}}]{Giallongo2015}
{Giallongo}, E., {Grazian}, A., {Fiore}, F., {et~al.} 2015, \aap, 578, A83,
  \dodoi{10.1051/0004-6361/201425334}

\bibitem[{{Gilfanov} \& {Merloni}(2014)}]{Gilfanov2014}
{Gilfanov}, M., \& {Merloni}, A. 2014, \ssr, 183, 121,
  \dodoi{10.1007/s11214-014-0071-5}

\bibitem[{{Greene} \& {Ho}(2005)}]{Greene2005}
{Greene}, J.~E., \& {Ho}, L.~C. 2005, \apj, 627, 721, \dodoi{10.1086/430590}

\bibitem[{{Grimm} {et~al.}(2003){Grimm}, {Gilfanov}, \& {Sunyaev}}]{Grimm2003}
{Grimm}, H.~J., {Gilfanov}, M., \& {Sunyaev}, R. 2003, \mnras, 339, 793,
  \dodoi{10.1046/j.1365-8711.2003.06224.x}

\bibitem[{Groves {et~al.}(2006)Groves, Heckman, \& Kauffmann}]{Groves2006}
Groves, B.~A., Heckman, T.~M., \& Kauffmann, G. 2006, Monthly Notices of the
  Royal Astronomical Society, \dodoi{10.1111/j.1365-2966.2006.10812.x}

\bibitem[{{Habouzit}(2024)}]{Habouzit2024}
{Habouzit}, M. 2024, arXiv e-prints, arXiv:2405.05319,
  \dodoi{10.48550/arXiv.2405.05319}

\bibitem[{{Haines} {et~al.}(2012){Haines}, {Pereira}, {Sanderson}, {Smith},
  {Egami}, {Babul}, {Edge}, {Finoguenov}, {Moran}, \& {Okabe}}]{Haines2012}
{Haines}, C.~P., {Pereira}, M.~J., {Sanderson}, A.~J.~R., {et~al.} 2012, \apj,
  754, 97, \dodoi{10.1088/0004-637X/754/2/97}

\bibitem[{{Harikane} {et~al.}(2023){Harikane}, {Zhang}, {Nakajima}, {Ouchi},
  {Isobe}, {Ono}, {Hatano}, {Xu}, \& {Umeda}}]{Harikane2023}
{Harikane}, Y., {Zhang}, Y., {Nakajima}, K., {et~al.} 2023, arXiv e-prints,
  arXiv:2303.11946, \dodoi{10.48550/arXiv.2303.11946}

\bibitem[{{Harish} {et~al.}(2021){Harish}, {Malhotra}, {Rhoads}, {Jiang},
  {Yang}, \& {Knorr}}]{Harish2021}
{Harish}, S., {Malhotra}, S., {Rhoads}, J.~E., {et~al.} 2021, arXiv e-prints,
  arXiv:2105.13400, \dodoi{10.48550/arXiv.2105.13400}

\bibitem[{{Heckman} {et~al.}(1981){Heckman}, {Miley}, {van Breugel}, \&
  {Butcher}}]{Heckman1981}
{Heckman}, T.~M., {Miley}, G.~K., {van Breugel}, W.~J.~M., \& {Butcher}, H.~R.
  1981, \apj, 247, 403, \dodoi{10.1086/159050}

\bibitem[{{Henry} {et~al.}(2015){Henry}, {Scarlata}, {Martin}, \&
  {Erb}}]{Henry2015}
{Henry}, A., {Scarlata}, C., {Martin}, C.~L., \& {Erb}, D. 2015, \apj, 809, 19,
  \dodoi{10.1088/0004-637X/809/1/19}

\bibitem[{{Israel} {et~al.}(2017){Israel}, {Belfiore}, {Stella}, {Esposito},
  {Casella}, {De Luca}, {Marelli}, {Papitto}, {Perri}, {Puccetti}, {Castillo},
  {Salvetti}, {Tiengo}, {Zampieri}, {D'Agostino}, {Greiner}, {Haberl},
  {Novara}, {Salvaterra}, {Turolla}, {Watson}, {Wilms}, \&
  {Wolter}}]{Israel2017}
{Israel}, G.~L., {Belfiore}, A., {Stella}, L., {et~al.} 2017, Science, 355,
  817, \dodoi{10.1126/science.aai8635}

\bibitem[{{Izotov} {et~al.}(2011{\natexlab{a}}){Izotov}, {Guseva}, {Fricke}, \&
  {Henkel}}]{Izotov2011}
{Izotov}, Y.~I., {Guseva}, N.~G., {Fricke}, K.~J., \& {Henkel}, C.
  2011{\natexlab{a}}, \aap, 536, L7, \dodoi{10.1051/0004-6361/201118402}

\bibitem[{{Izotov} {et~al.}(2011{\natexlab{b}}){Izotov}, {Guseva}, \&
  {Thuan}}]{Izotov2011a}
{Izotov}, Y.~I., {Guseva}, N.~G., \& {Thuan}, T.~X. 2011{\natexlab{b}}, \apj,
  728, 161, \dodoi{10.1088/0004-637X/728/2/161}

\bibitem[{{Jarrett} {et~al.}(2011){Jarrett}, {Cohen}, {Masci}, {Wright},
  {Stern}, {Benford}, {Blain}, {Carey}, {Cutri}, {Eisenhardt}, {Lonsdale},
  {Mainzer}, {Marsh}, {Padgett}, {Petty}, {Ressler}, {Skrutskie}, {Stanford},
  {Surace}, {Tsai}, {Wheelock}, \& {Yan}}]{Jarrett2011}
{Jarrett}, T.~H., {Cohen}, M., {Masci}, F., {et~al.} 2011, \apj, 735, 112,
  \dodoi{10.1088/0004-637X/735/2/112}

\bibitem[{{Jiang} {et~al.}(2019){Jiang}, {Malhotra}, {Yang}, \&
  {Rhoads}}]{Jiang2019}
{Jiang}, T., {Malhotra}, S., {Yang}, H., \& {Rhoads}, J.~E. 2019, \apj, 872,
  146, \dodoi{10.3847/1538-4357/aaee79}

\bibitem[{{Juod{\v{z}}balis} {et~al.}(2023){Juod{\v{z}}balis}, {Conselice},
  {Singh}, {Adams}, {Ormerod}, {Harvey}, {Austin}, {Volonteri}, {Cohen},
  {Jansen}, {Summers}, {Windhorst}, {D'Silva}, {Koekemoer}, {Coe}, {Driver},
  {Frye}, {Grogin}, {Marshall}, {Nonino}, {Pirzkal}, {Robotham}, {}, {Ortiz},
  {Tompkins}, {Willmer}, \& {Yan}}]{Juodzbalis2023}
{Juod{\v{z}}balis}, I., {Conselice}, C.~J., {Singh}, M., {et~al.} 2023, \mnras,
  525, 1353, \dodoi{10.1093/mnras/stad2396}

\bibitem[{{Kaaret} {et~al.}(2017){Kaaret}, {Feng}, \& {Roberts}}]{Kaaret2017}
{Kaaret}, P., {Feng}, H., \& {Roberts}, T.~P. 2017, \araa, 55, 303,
  \dodoi{10.1146/annurev-astro-091916-055259}

\bibitem[{{Karino} \& {Miller}(2016)}]{Karino2016}
{Karino}, S., \& {Miller}, J.~C. 2016, \mnras, 462, 3476,
  \dodoi{10.1093/mnras/stw1180}

\bibitem[{{Kauffmann} {et~al.}(2003){Kauffmann}, {Heckman}, {White}, {Charlot},
  {Tremonti}, {Brinchmann}, {Bruzual}, {Peng}, {Seibert}, {Bernardi},
  {Blanton}, {Brinkmann}, {Castander}, {Cs{\'a}bai}, {Fukugita}, {Ivezic},
  {Munn}, {Nichol}, {Padmanabhan}, {Thakar}, {Weinberg}, \&
  {York}}]{Kauffmann2003}
{Kauffmann}, G., {Heckman}, T.~M., {White}, S. D.~M., {et~al.} 2003, \mnras,
  341, 33, \dodoi{10.1046/j.1365-8711.2003.06291.x}

\bibitem[{{Kewley} {et~al.}(2001){Kewley}, {Dopita}, {Sutherland}, {Heisler},
  \& {Trevena}}]{Kewley2001}
{Kewley}, L.~J., {Dopita}, M.~A., {Sutherland}, R.~S., {Heisler}, C.~A., \&
  {Trevena}, J. 2001, \apj, 556, 121, \dodoi{10.1086/321545}

\bibitem[{{Kim} {et~al.}(2021){Kim}, {Malhotra}, {Rhoads}, \& {Yang}}]{Kim2021}
{Kim}, K.~J., {Malhotra}, S., {Rhoads}, J.~E., \& {Yang}, H. 2021, \apj, 914,
  2, \dodoi{10.3847/1538-4357/abf833}

\bibitem[{{Kocevski} {et~al.}(2023){Kocevski}, {Onoue}, {Inayoshi}, {Trump},
  {Arrabal Haro}, {Grazian}, {Dickinson}, {Finkelstein}, {Kartaltepe},
  {Hirschmann}, {Aird}, {Holwerda}, {Fujimoto}, {Juneau}, {Amor{\'\i}n},
  {Backhaus}, {Bagley}, {Barro}, {Bell}, {Bisigello}, {Calabr{\`o}}, {Cleri},
  {Cooper}, {Ding}, {Grogin}, {Ho}, {Hutchison}, {Inoue}, {Jiang}, {Jones},
  {Koekemoer}, {Li}, {Li}, {McGrath}, {Molina}, {Papovich},
  {P{\'e}rez-Gonz{\'a}lez}, {Pirzkal}, {Wilkins}, {Yang}, \&
  {Yung}}]{Kocevski2023}
{Kocevski}, D.~D., {Onoue}, M., {Inayoshi}, K., {et~al.} 2023, \apjl, 954, L4,
  \dodoi{10.3847/2041-8213/ace5a0}

\bibitem[{{Kocevski} {et~al.}(2024){Kocevski}, {Finkelstein}, {Barro},
  {Taylor}, {Calabr{\`o}}, {Laloux}, {Buchner}, {Trump}, {Leung}, {Yang},
  {Dickinson}, {P{\'e}rez-Gonz{\'a}lez}, {Pacucci}, {Inayoshi}, {Somerville},
  {McGrath}, {Akins}, {Bagley}, {Bisigello}, {Bowler}, {Carnall}, {Casey},
  {Cheng}, {Cleri}, {Costantin}, {Cullen}, {Davis}, {Donnan}, {Dunlop},
  {Ellis}, {Ferguson}, {Fujimoto}, {Fontana}, {Giavalisco}, {Grazian},
  {Grogin}, {Hathi}, {Hirschmann}, {Huertas-Company}, {Holwerda},
  {Illingworth}, {Juneau}, {Kartaltepe}, {Koekemoer}, {Li}, {Lucas}, {Magee},
  {Mason}, {McLeod}, {McLure}, {Napolitano}, {Papovich}, {Pirzkal},
  {Rodighiero}, {Santini}, {Wilkins}, \& {Yung}}]{Kocevski2024}
{Kocevski}, D.~D., {Finkelstein}, S.~L., {Barro}, G., {et~al.} 2024, arXiv
  e-prints, arXiv:2404.03576, \dodoi{10.48550/arXiv.2404.03576}

\bibitem[{{Kubota} \& {Done}(2018)}]{Kubota2018}
{Kubota}, A., \& {Done}, C. 2018, \mnras, 480, 1247,
  \dodoi{10.1093/mnras/sty1890}

\bibitem[{{Larson} {et~al.}(2023){Larson}, {Finkelstein}, {Kocevski},
  {Hutchison}, {Trump}, {Arrabal Haro}, {Bromm}, {Cleri}, {Dickinson},
  {Fujimoto}, {Kartaltepe}, {Koekemoer}, {Papovich}, {Pirzkal}, {Tacchella},
  {Zavala}, {Bagley}, {Behroozi}, {Champagne}, {Cole}, {Jung}, {Morales},
  {Yang}, {Zhang}, {Zitrin}, {Amor{\'\i}n}, {Burgarella}, {Casey}, {Ch{\'a}vez
  Ortiz}, {Cox}, {Chworowsky}, {Fontana}, {Gawiser}, {Grazian}, {Grogin},
  {Harish}, {Hathi}, {Hirschmann}, {Holwerda}, {Juneau}, {Leung}, {Lucas},
  {McGrath}, {P{\'e}rez-Gonz{\'a}lez}, {Rigby}, {Seill{\'e}}, {Simons}, {de La
  Vega}, {Weiner}, {Wilkins}, {Yung}, \& {Ceers Team}}]{Larson2023}
{Larson}, R.~L., {Finkelstein}, S.~L., {Kocevski}, D.~D., {et~al.} 2023, \apjl,
  953, L29, \dodoi{10.3847/2041-8213/ace619}

\bibitem[{{Latimer} {et~al.}(2021){Latimer}, {Reines}, {Bogdan}, \&
  {Kraft}}]{Latimer2021}
{Latimer}, L.~J., {Reines}, A.~E., {Bogdan}, A., \& {Kraft}, R. 2021, \apjl,
  922, L40, \dodoi{10.3847/2041-8213/ac3af6}

\bibitem[{{Lehmer} {et~al.}(2010){Lehmer}, {Alexander}, {Bauer}, {Brandt},
  {Goulding}, {Jenkins}, {Ptak}, \& {Roberts}}]{Lehmer2010}
{Lehmer}, B.~D., {Alexander}, D.~M., {Bauer}, F.~E., {et~al.} 2010, \apj, 724,
  559, \dodoi{10.1088/0004-637X/724/1/559}

\bibitem[{{Lehmer} {et~al.}(2022){Lehmer}, {Eufrasio}, {Basu-Zych}, {Garofali},
  {Gilbertson}, {Mesinger}, \& {Yukita}}]{Lehmer2022}
{Lehmer}, B.~D., {Eufrasio}, R.~T., {Basu-Zych}, A., {et~al.} 2022, \apj, 930,
  135, \dodoi{10.3847/1538-4357/ac63a7}

\bibitem[{{Lehmer} {et~al.}(2019){Lehmer}, {Eufrasio}, {Tzanavaris},
  {Basu-Zych}, {Fragos}, {Prestwich}, {Yukita}, {Zezas}, {Hornschemeier}, \&
  {Ptak}}]{Lehmer2019}
{Lehmer}, B.~D., {Eufrasio}, R.~T., {Tzanavaris}, P., {et~al.} 2019, \apjs,
  243, 3, \dodoi{10.3847/1538-4365/ab22a8}

\bibitem[{{Lehmer} {et~al.}(2021){Lehmer}, {Eufrasio}, {Basu-Zych}, {Doore},
  {Fragos}, {Garofali}, {Kovlakas}, {Williams}, {Zezas}, \&
  {Santana-Silva}}]{Lehmer2021}
{Lehmer}, B.~D., {Eufrasio}, R.~T., {Basu-Zych}, A., {et~al.} 2021, \apj, 907,
  17, \dodoi{10.3847/1538-4357/abcec1}

\bibitem[{{Lin} {et~al.}(2018){Lin}, {Strader}, {Carrasco}, {Page},
  {Romanowsky}, {Homan}, {Irwin}, {Remillard}, {Godet}, {Webb}, {Baumgardt},
  {Wijnands}, {Barret}, {Duc}, {Brodie}, \& {Gwyn}}]{Lin2018}
{Lin}, D., {Strader}, J., {Carrasco}, E.~R., {et~al.} 2018, Nature Astronomy,
  2, 656, \dodoi{10.1038/s41550-018-0493-1}

\bibitem[{{Madau} \& {Haardt}(2015)}]{Madau2015}
{Madau}, P., \& {Haardt}, F. 2015, \apjl, 813, L8,
  \dodoi{10.1088/2041-8205/813/1/L8}

\bibitem[{{Malhotra} {et~al.}(2012){Malhotra}, {Rhoads}, {Finkelstein},
  {Hathi}, {Nilsson}, {McLinden}, \& {Pirzkal}}]{Malhotra2012}
{Malhotra}, S., {Rhoads}, J.~E., {Finkelstein}, S.~L., {et~al.} 2012, \apjl,
  750, L36, \dodoi{10.1088/2041-8205/750/2/L36}

\bibitem[{{Malhotra} {et~al.}(2003){Malhotra}, {Wang}, {Rhoads}, {Heckman}, \&
  {Norman}}]{Malhotra2003}
{Malhotra}, S., {Wang}, J.~X., {Rhoads}, J.~E., {Heckman}, T.~M., \& {Norman},
  C.~A. 2003, \apjl, 585, L25, \dodoi{10.1086/373917}

\bibitem[{{Marin} {et~al.}(2023){Marin}, {Churazov}, {Khabibullin},
  {Ferrazzoli}, {Di Gesu}, {Barnouin}, {Di Marco}, {Middei}, {Vikhlinin},
  {Costa}, {Soffitta}, {Muleri}, {Sunyaev}, {Forman}, {Kraft}, {Bianchi},
  {Donnarumma}, {Petrucci}, {Enoto}, {Agudo}, {Antonelli}, {Bachetti},
  {Baldini}, {Baumgartner}, {Bellazzini}, {Bongiorno}, {Bonino}, {Brez},
  {Bucciantini}, {Capitanio}, {Castellano}, {Cavazzuti}, {Chen}, {Ciprini}, {De
  Rosa}, {Del Monte}, {Di Lalla}, {Doroshenko}, {Dov{\v{c}}iak}, {Ehlert},
  {Evangelista}, {Fabiani}, {Garcia}, {Gunji}, {Hayashida}, {Heyl}, {Ingram},
  {Iwakiri}, {Jorstad}, {Kaaret}, {Karas}, {Kitaguchi}, {Kolodziejczak},
  {Krawczynski}, {La Monaca}, {Latronico}, {Liodakis}, {Maldera}, {Manfreda},
  {Marinucci}, {Marscher}, {Marshall}, {Massaro}, {Matt}, {Mitsuishi},
  {Mizuno}, {Negro}, {Ng}, {O'Dell}, {Omodei}, {Oppedisano}, {Papitto},
  {Pavlov}, {Peirson}, {Perri}, {Pesce-Rollins}, {Pilia}, {Possenti},
  {Poutanen}, {Puccetti}, {Ramsey}, {Rankin}, {Ratheesh}, {Roberts}, {Romani},
  {Sgr{\`o}}, {Slane}, {Spandre}, {Swartz}, {Tamagawa}, {Tavecchio}, {Taverna},
  {Tawara}, {Tennant}, {Thomas}, {Tombesi}, {Trois}, {Tsygankov}, {Turolla},
  {Vink}, {Weisskopf}, {Wu}, {Xie}, \& {Zane}}]{Marin2023}
{Marin}, F., {Churazov}, E., {Khabibullin}, I., {et~al.} 2023, \nat, 619, 41,
  \dodoi{10.1038/s41586-023-06064-x}

\bibitem[{{Mezcua} {et~al.}(2018){Mezcua}, {Civano}, {Marchesi}, {Suh},
  {Fabbiano}, \& {Volonteri}}]{Mezcua2018}
{Mezcua}, M., {Civano}, F., {Marchesi}, S., {et~al.} 2018, \mnras, 478, 2576,
  \dodoi{10.1093/mnras/sty1163}

\bibitem[{{Mineo} {et~al.}(2014){Mineo}, {Gilfanov}, {Lehmer}, {Morrison}, \&
  {Sunyaev}}]{Mineo2014}
{Mineo}, S., {Gilfanov}, M., {Lehmer}, B.~D., {Morrison}, G.~E., \& {Sunyaev},
  R. 2014, \mnras, 437, 1698, \dodoi{10.1093/mnras/stt1999}

\bibitem[{{Mineo} {et~al.}(2012){Mineo}, {Gilfanov}, \& {Sunyaev}}]{Mineo2012}
{Mineo}, S., {Gilfanov}, M., \& {Sunyaev}, R. 2012, \mnras, 426, 1870,
  \dodoi{10.1111/j.1365-2966.2012.21831.x}

\bibitem[{{Naidu} {et~al.}(2020){Naidu}, {Tacchella}, {Mason}, {Bose}, {Oesch},
  \& {Conroy}}]{Naidu2020}
{Naidu}, R.~P., {Tacchella}, S., {Mason}, C.~A., {et~al.} 2020, \apj, 892, 109,
  \dodoi{10.3847/1538-4357/ab7cc9}

\bibitem[{{Nakajima} {et~al.}(2023){Nakajima}, {Ouchi}, {Isobe}, {Harikane},
  {Zhang}, {Ono}, {Umeda}, \& {Oguri}}]{Nakajima2023}
{Nakajima}, K., {Ouchi}, M., {Isobe}, Y., {et~al.} 2023, arXiv e-prints,
  arXiv:2301.12825, \dodoi{10.48550/arXiv.2301.12825}

\bibitem[{{Nandra} {et~al.}(2002){Nandra}, {Mushotzky}, {Arnaud}, {Steidel},
  {Adelberger}, {Gardner}, {Teplitz}, \& {Windhorst}}]{Nandra2002}
{Nandra}, K., {Mushotzky}, R.~F., {Arnaud}, K., {et~al.} 2002, \apj, 576, 625,
  \dodoi{10.1086/341888}

\bibitem[{{Ohsuga} {et~al.}(2005){Ohsuga}, {Mori}, {Nakamoto}, \&
  {Mineshige}}]{Ohsuga2005}
{Ohsuga}, K., {Mori}, M., {Nakamoto}, T., \& {Mineshige}, S. 2005, \apj, 628,
  368, \dodoi{10.1086/430728}

\bibitem[{{Pintore} {et~al.}(2018){Pintore}, {Belfiore}, {Novara},
  {Salvaterra}, {Marelli}, {De Luca}, {Rigoselli}, {Israel}, {Rodriguez},
  {Mereghetti}, {Wolter}, {Walton}, {Fuerst}, {Ambrosi}, {Zampieri}, {Tiengo},
  \& {Salvaggio}}]{Pintore2018}
{Pintore}, F., {Belfiore}, A., {Novara}, G., {et~al.} 2018, \mnras, 477, L90,
  \dodoi{10.1093/mnrasl/sly048}

\bibitem[{{Raassen} {et~al.}(2003){Raassen}, {van der Hucht}, {Mewe},
  {Antokhin}, {Rauw}, {Vreux}, {Schmutz}, \& {G{\"u}del}}]{Raaassen2003}
{Raassen}, A.~J.~J., {van der Hucht}, K.~A., {Mewe}, R., {et~al.} 2003, \aap,
  402, 653, \dodoi{10.1051/0004-6361:20030119}

\bibitem[{{Ranalli} {et~al.}(2003){Ranalli}, {Comastri}, \&
  {Setti}}]{Ranalli2003}
{Ranalli}, P., {Comastri}, A., \& {Setti}, G. 2003, \aap, 399, 39,
  \dodoi{10.1051/0004-6361:20021600}

\bibitem[{{Reines} {et~al.}(2014){Reines}, {Plotkin}, {Russell}, {Mezcua},
  {Condon}, {Sivakoff}, \& {Johnson}}]{Reines2014}
{Reines}, A.~E., {Plotkin}, R.~M., {Russell}, T.~D., {et~al.} 2014, \apjl, 787,
  L30, \dodoi{10.1088/2041-8205/787/2/L30}

\bibitem[{{Rephaeli} {et~al.}(1995){Rephaeli}, {Gruber}, \&
  {Persic}}]{Rephaeli1995}
{Rephaeli}, Y., {Gruber}, D., \& {Persic}, M. 1995, \aap, 300, 91

\bibitem[{{Rhoads} {et~al.}(2023){Rhoads}, {Wold}, {Harish}, {Kim}, {Pharo},
  {Malhotra}, {Gabrielpillai}, {Jiang}, \& {Yang}}]{Rhoads2023}
{Rhoads}, J.~E., {Wold}, I. G.~B., {Harish}, S., {et~al.} 2023, \apjl, 942,
  L14, \dodoi{10.3847/2041-8213/acaaaf}

\bibitem[{{Ricci} {et~al.}(2020){Ricci}, {Kara}, {Loewenstein}, {Trakhtenbrot},
  {Arcavi}, {Remillard}, {Fabian}, {Gendreau}, {Arzoumanian}, {Li}, {Ho},
  {MacLeod}, {Cackett}, {Altamirano}, {Gandhi}, {Kosec}, {Pasham}, {Steiner},
  \& {Chan}}]{Ricci2020}
{Ricci}, C., {Kara}, E., {Loewenstein}, M., {et~al.} 2020, \apjl, 898, L1,
  \dodoi{10.3847/2041-8213/ab91a1}

\bibitem[{{Robertson} {et~al.}(2015){Robertson}, {Ellis}, {Furlanetto}, \&
  {Dunlop}}]{Robertson2015}
{Robertson}, B.~E., {Ellis}, R.~S., {Furlanetto}, S.~R., \& {Dunlop}, J.~S.
  2015, \apjl, 802, L19, \dodoi{10.1088/2041-8205/802/2/L19}

\bibitem[{{Sabha} {et~al.}(2010){Sabha}, {Witzel}, {Eckart}, {Buchholz},
  {Bremer}, {Gie{\ss}{\"u}bel}, {Garc{\'\i}a-Mar{\'\i}n}, {Kunneriath},
  {Muzic}, {Sch{\"o}del}, {Straubmeier}, {Zamaninasab}, \&
  {Zernickel}}]{Sabha2010}
{Sabha}, N., {Witzel}, G., {Eckart}, A., {et~al.} 2010, \aap, 512, A2,
  \dodoi{10.1051/0004-6361/200913186}

\bibitem[{{Satyapal} {et~al.}(2018){Satyapal}, {Abel}, \&
  {Secrest}}]{Satyapal2018}
{Satyapal}, S., {Abel}, N.~P., \& {Secrest}, N.~J. 2018, \apj, 858, 38,
  \dodoi{10.3847/1538-4357/aab7f8}

\bibitem[{{Satyapal} {et~al.}(2021){Satyapal}, {Kamal}, {Cann}, {Secrest}, \&
  {Abel}}]{Satyapal2021}
{Satyapal}, S., {Kamal}, L., {Cann}, J.~M., {Secrest}, N.~J., \& {Abel}, N.~P.
  2021, \apj, 906, 35, \dodoi{10.3847/1538-4357/abbfaf}

\bibitem[{{Shakura} \& {Sunyaev}(1973)}]{Shakura1973}
{Shakura}, N.~I., \& {Sunyaev}, R.~A. 1973, \aap, 24, 337

\bibitem[{{She} {et~al.}(2018){She}, {Ho}, {Feng}, \& {Cui}}]{She2018}
{She}, R., {Ho}, L.~C., {Feng}, H., \& {Cui}, C. 2018, \apj, 859, 152,
  \dodoi{10.3847/1538-4357/aabfe7}

\bibitem[{{Singha} {et~al.}(2021){Singha}, {O'Dea}, {Gordon}, {Lawlor-Forsyth},
  \& {Baum}}]{Singha2021}
{Singha}, M., {O'Dea}, C.~P., {Gordon}, Y.~A., {Lawlor-Forsyth}, C., \& {Baum},
  S.~A. 2021, \apj, 918, 65, \dodoi{10.3847/1538-4357/ac06c7}

\bibitem[{{Singha} {et~al.}(2022){Singha}, {Husemann}, {Urrutia}, {O'Dea},
  {Scharw{\"a}chter}, {Gaspari}, {Combes}, {Nevin}, {Terrazas},
  {P{\'e}rez-Torres}, {Rose}, {Davis}, {Tremblay}, {Neumann},
  {Smirnova-Pinchukova}, \& {Baum}}]{Singha2022}
{Singha}, M., {Husemann}, B., {Urrutia}, T., {et~al.} 2022, \aap, 659, A123,
  \dodoi{10.1051/0004-6361/202040122}

\bibitem[{{Singha} {et~al.}(2023){Singha}, {Winkel}, {Vaddi}, {Perez Torres},
  {Gaspari}, {Smirnova-Pinchukova}, {O'Dea}, {Combes}, {Omoruyi}, {Rose},
  {McElroy}, {Husemann}, {Davis}, {Baum}, {Lawlor-Forsyth}, {Neumann}, \&
  {Tremblay}}]{Singha2023}
{Singha}, M., {Winkel}, N., {Vaddi}, S., {et~al.} 2023, \apj, 959, 107,
  \dodoi{10.3847/1538-4357/ad004d}

\bibitem[{{Sutton} {et~al.}(2013){Sutton}, {Roberts}, {Gladstone}, {Farrell},
  {Reilly}, {Goad}, \& {Gehrels}}]{Sutton2013}
{Sutton}, A.~D., {Roberts}, T.~P., {Gladstone}, J.~C., {et~al.} 2013, \mnras,
  434, 1702, \dodoi{10.1093/mnras/stt1133}

\bibitem[{{Svoboda} {et~al.}(2019){Svoboda}, {Douna}, {Orlitov{\'a}}, \&
  {Ehle}}]{Svoboda2019}
{Svoboda}, J., {Douna}, V., {Orlitov{\'a}}, I., \& {Ehle}, M. 2019, \apj, 880,
  144, \dodoi{10.3847/1538-4357/ab2b39}

\bibitem[{{Swartz} {et~al.}(2011){Swartz}, {Soria}, {Tennant}, \&
  {Yukita}}]{Swartz2011}
{Swartz}, D.~A., {Soria}, R., {Tennant}, A.~F., \& {Yukita}, M. 2011, \apj,
  741, 49, \dodoi{10.1088/0004-637X/741/1/49}

\bibitem[{{Volonteri} {et~al.}(2016){Volonteri}, {Dubois}, {Pichon}, \&
  {Devriendt}}]{Volonteri2016}
{Volonteri}, M., {Dubois}, Y., {Pichon}, C., \& {Devriendt}, J. 2016, \mnras,
  460, 2979, \dodoi{10.1093/mnras/stw1123}

\bibitem[{{Wang} {et~al.}(2004){Wang}, {Malhotra}, {Rhoads}, \&
  {Norman}}]{Wang2004b}
{Wang}, J.~X., {Malhotra}, S., {Rhoads}, J.~E., \& {Norman}, C.~A. 2004, \apjl,
  612, L109, \dodoi{10.1086/424799}

\bibitem[{{White} \& {Long}(1986)}]{White1986}
{White}, R.~L., \& {Long}, K.~S. 1986, \apj, 310, 832, \dodoi{10.1086/164736}

\bibitem[{{Winkel} {et~al.}(2023){Winkel}, {Husemann}, {Singha}, {Bennert},
  {Combes}, {Davis}, {Gaspari}, {Jahnke}, {McElroy}, {O'Dea}, \&
  {P{\'e}rez-Torres}}]{Winkel2023}
{Winkel}, N., {Husemann}, B., {Singha}, M., {et~al.} 2023, \aap, 670, A3,
  \dodoi{10.1051/0004-6361/202244630}

\bibitem[{{Wood} {et~al.}(2015){Wood}, {Tremonti}, {Calzetti}, {Leitherer},
  {Chisholm}, \& {Gallagher}}]{Wood2015}
{Wood}, C.~M., {Tremonti}, C.~A., {Calzetti}, D., {et~al.} 2015, \mnras, 452,
  2712, \dodoi{10.1093/mnras/stv1471}

\bibitem[{{Wright} {et~al.}(2010){Wright}, {Eisenhardt}, {Mainzer}, {Ressler},
  {Cutri}, {Jarrett}, {Kirkpatrick}, {Padgett}, {McMillan}, {Skrutskie},
  {Stanford}, {Cohen}, {Walker}, {Mather}, {Leisawitz}, {Gautier}, {McLean},
  {Benford}, {Lonsdale}, {Blain}, {Mendez}, {Irace}, {Duval}, {Liu}, {Royer},
  {Heinrichsen}, {Howard}, {Shannon}, {Kendall}, {Walsh}, {Larsen}, {Cardon},
  {Schick}, {Schwalm}, {Abid}, {Fabinsky}, {Naes}, \& {Tsai}}]{Wright2010}
{Wright}, E.~L., {Eisenhardt}, P. R.~M., {Mainzer}, A.~K., {et~al.} 2010, \aj,
  140, 1868, \dodoi{10.1088/0004-6256/140/6/1868}

\bibitem[{{Xu} {et~al.}(2023){Xu}, {Ouchi}, {Nakajima}, {Harikane}, {Isobe},
  {Ono}, {Umeda}, \& {Zhang}}]{Xu2023}
{Xu}, Y., {Ouchi}, M., {Nakajima}, K., {et~al.} 2023, arXiv e-prints,
  arXiv:2310.06614, \dodoi{10.48550/arXiv.2310.06614}

\bibitem[{{Yang} {et~al.}(2023){Yang}, {Caputi}, {Papovich}, {Arrabal Haro},
  {Bagley}, {Behroozi}, {Bell}, {Bisigello}, {Buat}, {Burgarella}, {Cheng},
  {Cleri}, {Dav{\'e}}, {Dickinson}, {Elbaz}, {Ferguson}, {Finkelstein},
  {Grogin}, {Hathi}, {Hirschmann}, {Holwerda}, {Huertas-Company}, {Hutchison},
  {Iani}, {Kartaltepe}, {Kirkpatrick}, {Kocevski}, {Koekemoer}, {Kokorev},
  {Larson}, {Lucas}, {P{\'e}rez-Gonz{\'a}lez}, {Rinaldi}, {Shen}, {Trump}, {de
  la Vega}, {Yung}, \& {Zavala}}]{Yang2023}
{Yang}, G., {Caputi}, K.~I., {Papovich}, C., {et~al.} 2023, \apjl, 950, L5,
  \dodoi{10.3847/2041-8213/acd639}

\bibitem[{{Yang} {et~al.}(2016){Yang}, {Malhotra}, {Gronke}, {Rhoads},
  {Dijkstra}, {Jaskot}, {Zheng}, \& {Wang}}]{Yang2016}
{Yang}, H., {Malhotra}, S., {Gronke}, M., {et~al.} 2016, \apj, 820, 130,
  \dodoi{10.3847/0004-637X/820/2/130}

\bibitem[{{Yang} {et~al.}(2009){Yang}, {Zabludoff}, {Tremonti}, {Eisenstein},
  \& {Dav{\'e}}}]{Yang2009}
{Yang}, Y., {Zabludoff}, A., {Tremonti}, C., {Eisenstein}, D., \& {Dav{\'e}},
  R. 2009, \apj, 693, 1579, \dodoi{10.1088/0004-637X/693/2/1579}

\bibitem[{{Yung} {et~al.}(2021){Yung}, {Somerville}, {Finkelstein},
  {Hirschmann}, {Dav{\'e}}, {Popping}, {Gardner}, \& {Venkatesan}}]{Yung2021}
{Yung}, L.~Y.~A., {Somerville}, R.~S., {Finkelstein}, S.~L., {et~al.} 2021,
  \mnras, 508, 2706, \dodoi{10.1093/mnras/stab2761}

\bibitem[{{Yung} {et~al.}(2020{\natexlab{a}}){Yung}, {Somerville},
  {Finkelstein}, {Popping}, {Dav{\'e}}, {Venkatesan}, {Behroozi}, \&
  {Ferguson}}]{Yung2020b}
---. 2020{\natexlab{a}}, \mnras, 496, 4574, \dodoi{10.1093/mnras/staa1800}

\bibitem[{{Yung} {et~al.}(2020{\natexlab{b}}){Yung}, {Somerville}, {Popping},
  \& {Finkelstein}}]{Yung2020a}
{Yung}, L.~Y.~A., {Somerville}, R.~S., {Popping}, G., \& {Finkelstein}, S.~L.
  2020{\natexlab{b}}, \mnras, 494, 1002, \dodoi{10.1093/mnras/staa714}

\bibitem[{{Zhang}(2021)}]{Zhang2021}
{Zhang}, X.-G. 2021, \mnras, 502, 2508, \dodoi{10.1093/mnras/stab185}

\bibitem[{{Zheng} {et~al.}(2012){Zheng}, {Malhotra}, {Wang}, {Rhoads},
  {Finkelstein}, {Gawiser}, {Gronwall}, {Guaita}, {Nilsson}, \&
  {Ciardullo}}]{Zheng2012}
{Zheng}, Z.-Y., {Malhotra}, S., {Wang}, J.-X., {et~al.} 2012, \apj, 746, 28,
  \dodoi{10.1088/0004-637X/746/1/28}

\bibitem[{{Zuo} {et~al.}(2024){Zuo}, {Guo}, {Sun}, {Yuan}, {Lira}, {Gu},
  {Edwards}, {Gupta}, {Kishore}, {Stevens}, {An}, {Cai}, {Feng}, {Ho},
  {Ili{\'c}}, {Kova{\v{c}}evi{\'c}}, {Li}, {Mezcua}, {Popovi{\'c}}, {Sun},
  {Tripathi}, {U.}, {Vince}, {Wang}, {Wang}, {Wang}, {Wu}, \&
  {Zheng}}]{Zuo2024}
{Zuo}, W., {Guo}, H., {Sun}, J., {et~al.} 2024, arXiv e-prints,
  arXiv:2405.11750, \dodoi{10.48550/arXiv.2405.11750}

\end{thebibliography}

\section{Appendix}

\subsection{He~II$\lambda4686$ spectra}
\begin{figure*}[h!]
\centering
 \includegraphics[width=\textwidth]{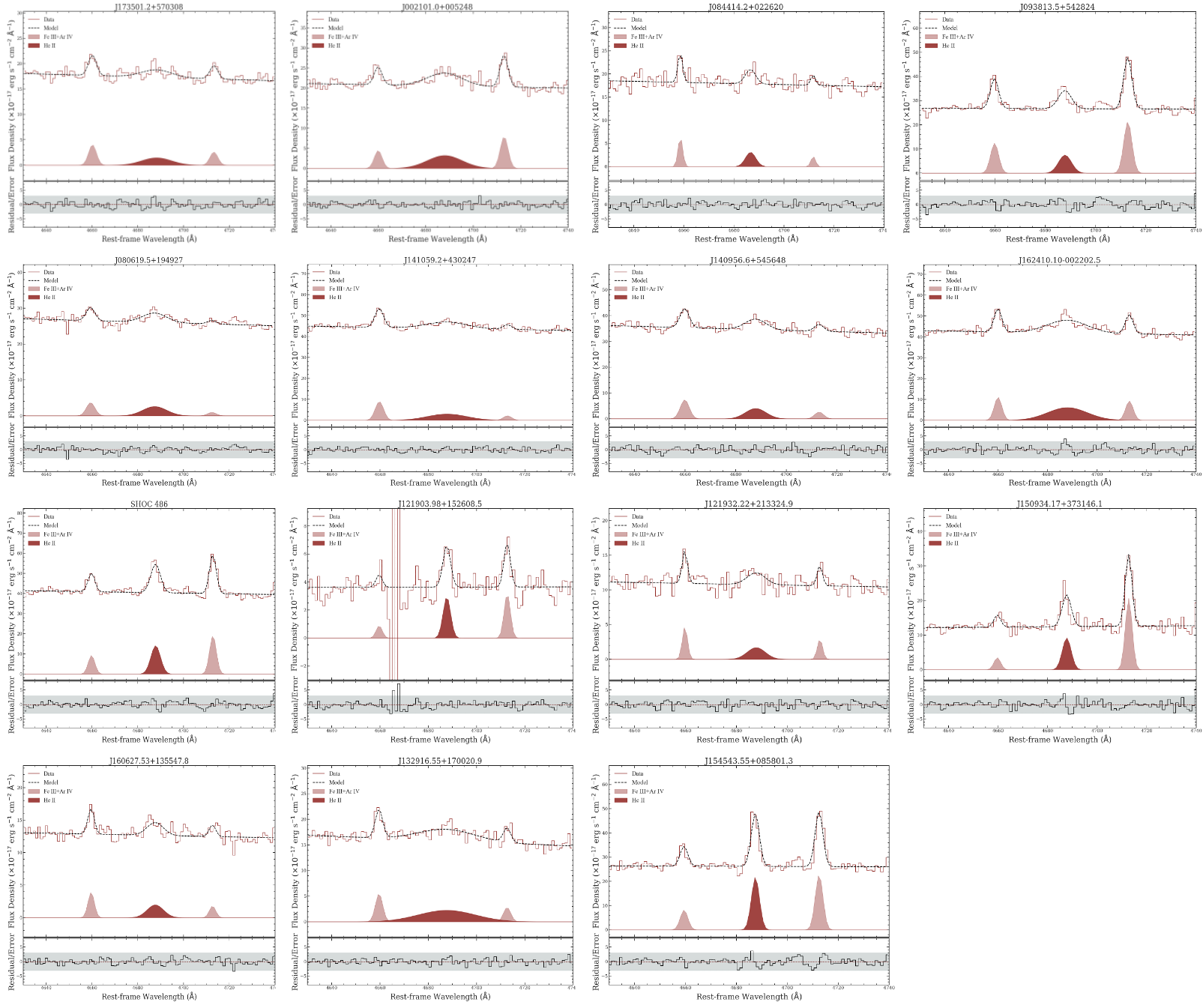}
 \caption{Top panel shows the source spectra centered on He~II$\lambda4686$. The solid red line is the data and the black dashed line is the composite model of the 3 emission lines which are present: Fe~III$\lambda4658$, He~II$\lambda4686$, and Ar~IV$\lambda4711$.  The shaded regions show the individual models for the corresponding lines.  The pink shaded region shows models for the Fe~III$\lambda4658$ and Ar~IV$\lambda4711$ lines. The red shaded regions shows the model for the He~II$\lambda4686$ line. Additional details regarding the models/fit can be found in Section \ref{sec:HeII emission}. The bottom panel shows residuals normalized by the error.}
\label{fig:fig_HeII_GP1}
\end{figure*}

\clearpage

\subsection{H$\alpha$+[N~II] spectra}

\begin{figure*}[h!]
\centering
 \includegraphics[width=\textwidth]{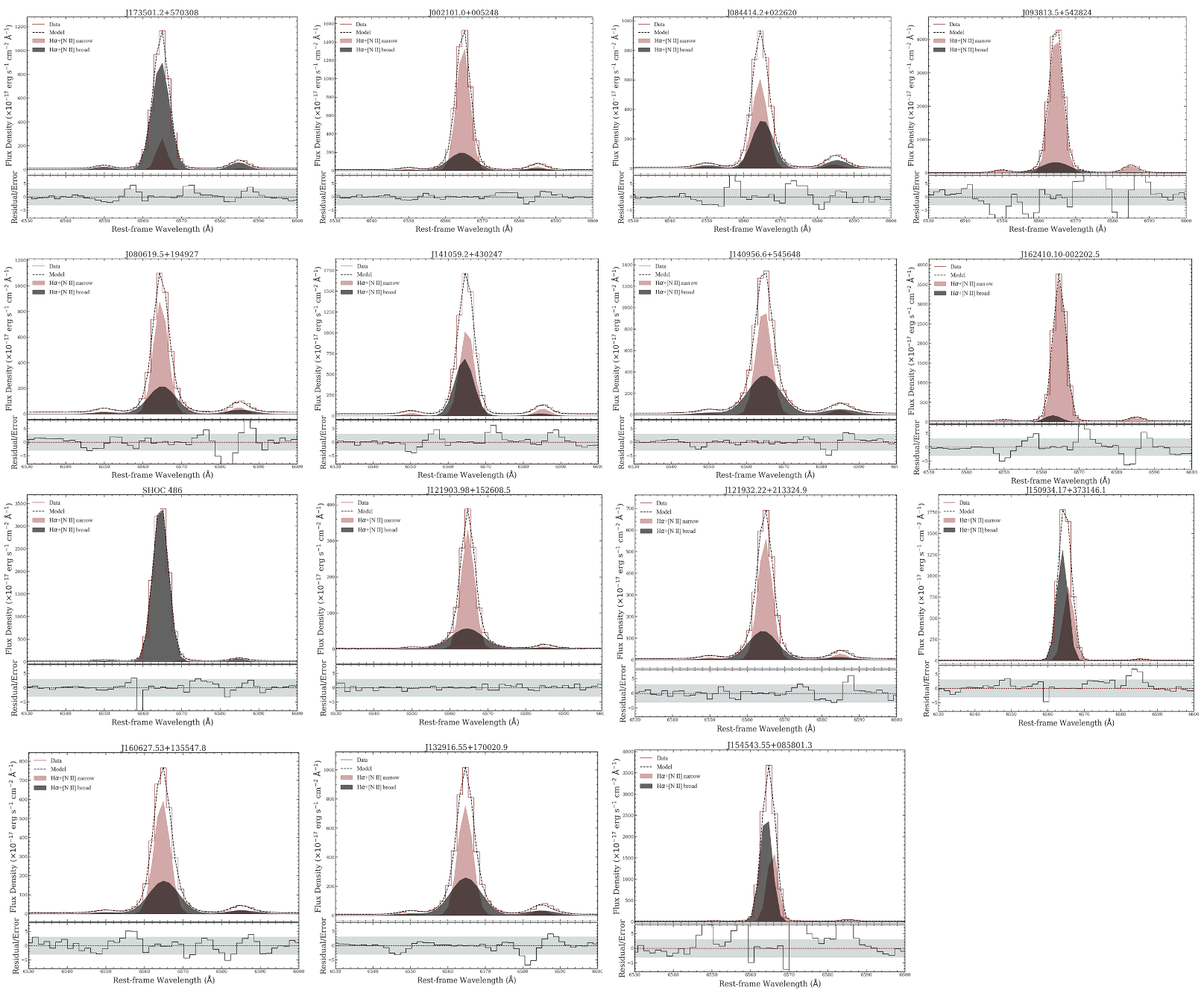}
 \caption{Top panel shows the source spectra centered on the H$\alpha$+[N~II]$\lambda\lambda6548,6584$ lines. The solid red line is the data and the black dashed line is the composite model of the emission lines. A multi-Gaussian model with both broad and narrow Gaussians was fit to the H$\alpha$ and [N II] lines.  The shaded regions show the components of the multi-Gaussian model.  The pink shaded region shows the narrow Gaussian model, while the dark grey shaded region shows the broad Gaussian model. Additional details regarding the models/fit can be found in Section \ref{sec:Ha emission}. The bottom panel shows residuals normalized by the error. }
\label{fig:fig_Ha2_GP1}
\end{figure*}

\clearpage

\subsection{H$\beta$+[O~III] spectra}

\begin{figure*}[h!]
\centering
 \includegraphics[width=\textwidth]{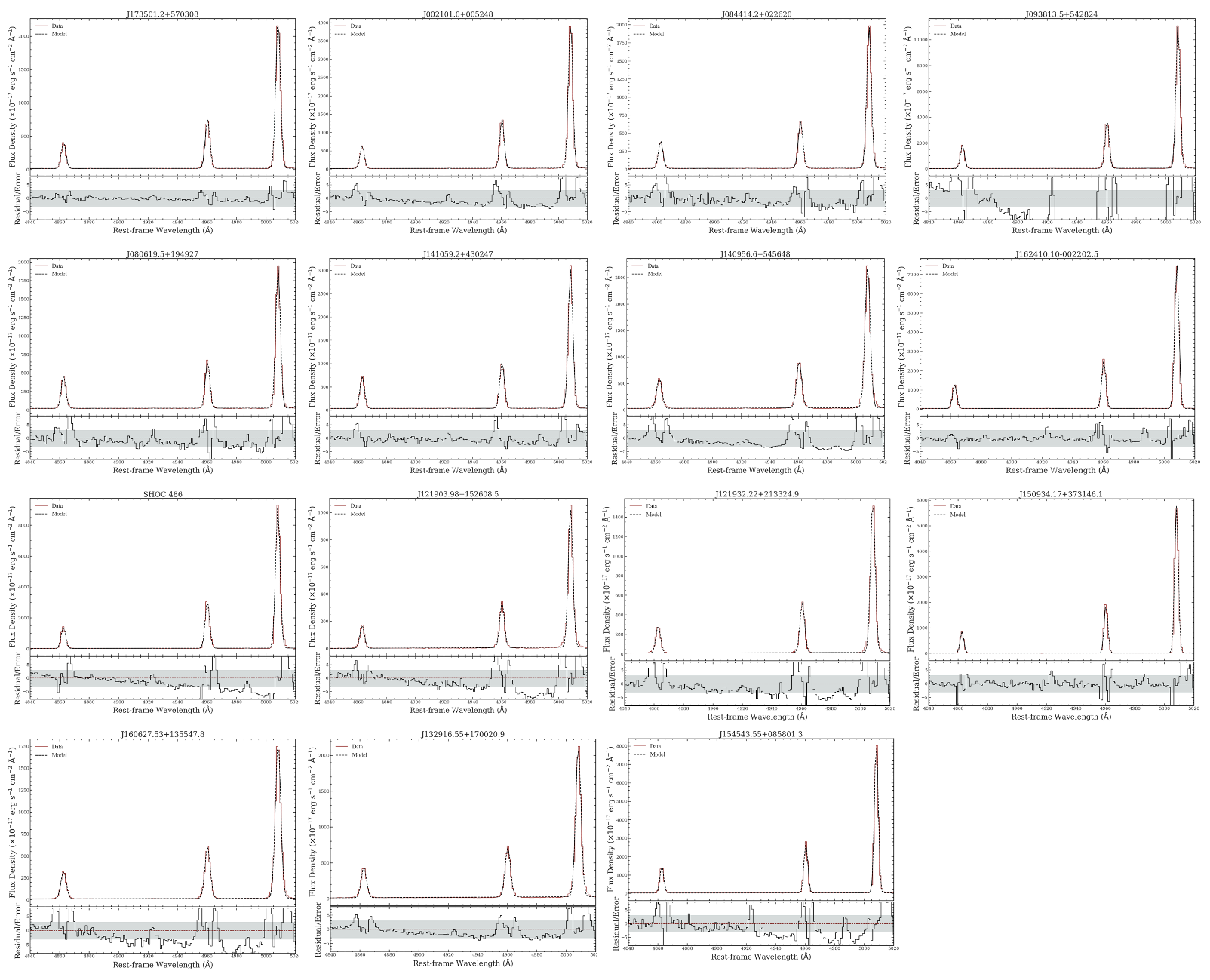}
 \caption{Top panel shows the source spectra centered on the H$\beta$+[O~III]$\lambda\lambda4959,5007$ lines. The solid red line is the data and the black dashed line is the composite model of the emission lines. A single Gaussian model for each of the aforementioned emission lines was fit to the emission line complex. The bottom panel shows residuals normalized by the error. Adding additional Gaussian did not significantly improve the fit.}
\label{fig:fig_Hb2_GP1}
\end{figure*}

\end{document}